\documentclass[12pt,a4paper,oneside]{article}

\usepackage{fourier}
\usepackage[T1]{fontenc}

\usepackage{amsfonts}
\usepackage{amssymb}
\usepackage{amsthm}
\usepackage{amsmath,graphicx,caption,mathtools}
\usepackage{relsize}
\usepackage{fullpage}
\usepackage{subcaption}
\usepackage{booktabs}
\usepackage{changepage}
\usepackage{natbib}
\usepackage[final, bookmarks, bookmarksnumbered=true, bookmarksopen=true, colorlinks=true, hyperfootnotes, linkcolor=black, anchorcolor=blue, citecolor=blue, filecolor=blue, menucolor=blue, runcolor=blue, urlcolor=black]{hyperref}
\usepackage{cleveref}
\usepackage{threeparttable}
\usepackage{tikz}
\usetikzlibrary{automata,positioning,fit, shapes.geometric,calc}
\usepackage{algorithm}
\usepackage[]{algpseudocode}
\usepackage{bbm}
\usepackage{array}
\usepackage{lscape}

\newcommand\blfootnote[1]{
  \begingroup
  \renewcommand\thefootnote{}\footnote{#1}
  \addtocounter{footnote}{-1}
  \endgroup
}

\newtheorem{theorem}{Theorem}
\newtheorem{lemma}{Lemma}
\newtheorem{assumption}{Assumption}
\newtheorem{corollary}{Corollary}
\newtheorem{proposition}{Proposition}

\begin{document}

\title{Instrument-based estimation of full treatment effects with movers 
}
\author{Didier Nibbering\thanks{Department of Econometrics and Business Statistics, Monash University, Melbourne, Australia. didier.nibbering@monash.edu} \and
Matthijs Oosterveen\thanks{Department of Economics, Lisbon School of Economics and Management, and Advance/CSG, University of Lisbon, Lisbon 1200-781, Portugal. oosterveen@iseg.ulisboa.pt} 
\blfootnote{
We thank Massimo Anelli, Akanksha Negi, Denni Tommasi, and Dinand Webbink for valuable comments. 
The authors have no relevant or material financial interests that relate to the research described in this paper. All omissions and errors are our own.
}
}
\date{\today \\ \vspace{0.2in}}

\maketitle
\thispagestyle{empty}

\vspace{-.3in}
\begin{abstract} 
\noindent {\normalsize 
The effect of the full treatment is a primary parameter of interest in policy evaluation, while often only the effect of a subset of treatment is estimated. We partially identify the local average treatment effect of receiving full treatment (LAFTE) using an instrumental variable that may induce individuals into only a subset of treatment (movers). We show that movers violate the standard exclusion restriction, necessary conditions on the presence of movers are testable, and partial identification holds under a double exclusion restriction. We identify movers in four empirical applications and estimate informative bounds on the LAFTE in three of them.
} 


\end{abstract}

\linespread{1.30}
\normalsize
\newpage
\setcounter{page}{1}

\section{Introduction}

This paper develops an instrumental variable (IV) framework that partially identifies the local average treatment effect (LATE) of receiving full treatment compared to no treatment. The effect of the full treatment program is a primary parameter of interest in policy evaluation. However, in many instances, policy evaluations yield estimates of only a subset of the treatment program. For instance, when the effect of college completion may be of interest, the effect of college enrollment is estimated. This is a problem for two main reasons. First, most treatment programs are designed to be received in full and the effects of separate treatment parts may be less informative. Second, there is often non-compliance with randomized treatment assignment and hence identification of treatment effects in an IV framework requires an exclusion restriction. This restriction may be violated by individuals moving in or out of treatment after the start of the program.

Individuals that move in or out of treatment are present in many policy evaluation settings. For instance, \cite{leu2017beginning} reports that approximately 30\% of college students changes major, \cite{sugar2021medicaid} describe the high prevalence of Medicaid coverage disruptions, often referred to as churning, due to income fluctuations, and \cite{heckman2000substitution} discuss the incidence of control group substitution and treatment group dropout across various job market training programs. This implies that instead of staying in or out of treatment during the whole program, there are late-adopters missing the first part of treatment or dropouts missing the final part of treatment. 

The IV exclusion restriction rules out certain types of movers depending on the definition of the treatment variable. For instance, \cite*{silliman2022labor}, \cite*{grosz2020returns}, \cite*{burde2013bringing}, and \cite*{hoekstra2009effect} take enrollment at the start of an educational program as treatment indicator. 
In this case, the exclusion restriction does not allow for movers that are induced by the instrument to only take the second part of the treatment. \cite{ketel2016returns} and \cite{zimmerman2014returns} use enrollment at the end of an educational program as treatment indicator. In this case, the exclusion restriction does not allow for movers that are induced by the instrument to only take the first part of the treatment. 

Researchers are familiar with the challenging restrictions imposed by the exclusion restriction. For instance, \cite{silliman2022labor} estimate the returns to vocational secondary education enrollment, and discuss how student dropout may violate the exclusion restriction. \cite{grosz2020returns} discusses how movers in and out of a nursing program affect the exclusion restriction with different definitions of treatment, such as enrolling at the start or ever enrolling, in the nursing program. \cite{finkelstein2012oregon} suggest that different definitions for the treatment variable of medicaid insurance can be used to provide a lower and upper bound for a LATE using the largest and smallest first stage, respectively.

This paper partially identifies the LATE of receiving full treatment compared to no treatment under a double exclusion restriction. This parameter is referred to as the local average full treatment effect (LAFTE). Our IV framework includes the binary potential treatment status for the first and second part of treatment with only one binary instrument. The instrument may induce individuals to obtain full treatment, or only the first or second part of the treatment. We refer to the latter group of individuals as movers. Within this framework, movers do not violate the IV exclusion restriction and LATEs are only identified under additional assumptions. The proposed double exclusion restriction extends this exclusion restriction --the instrument can only affect the outcome variable through treatment enrollment-- with the condition that the instrument can only affect treatment enrollment in the second part through enrollment in the first part. We provide a procedure for testing necessary conditions of the presence of movers and of the double exclusion restriction.

The partial identification of the LAFTE is achieved with nonparametric sharp bounds. The bounds formalize the intuition that different definitions for the treatment variable can be utilized to identify the effect of receiving a full treatment program, while allowing for movers. We show that the bounds can be tightened using the monotone treatment response and monotone treatment selection assumptions of \cite{manski1997monotone} and \cite{manski2000monotone} in addition to the double exclusion restriction. Less informative bounds rely solely on double exclusion and a bounded response. 

The double exclusion restriction holds if a delayed treatment response to treatment assignment is absent, as it rules out the movers that are induced by the instrument to only take the second part of treatment. It now follows that the common practice of using IV with enrollment in the first part of treatment as treatment indicator identifies the LATE of taking the first part of the treatment. Our double exclusion restriction is similar to, albeit weaker than, the dynamic exclusion restriction of \cite{angrist2022marginal}. The dynamic exclusion restriction imposes the additional condition that treatment enrollment in the first part can only affect the outcome variable though enrollment in the second part. This either rules out movers, restricts treatment effect heterogeneity, or a combination of these two. In this case, LAFTE can be point-identified using the IV method of \cite{imbens1994identification}.

The policy relevance of the LAFTE is illustrated with four empirical applications from different fields of applied economics: a health program by \citet{finkelstein2012oregon,finkelstein2016effect}, a labour program by \citet{wheeler2022linkedin}, an educational program by \citet{anelli2020returns}, and a development program by \citet{burde2013bringing}. For instance, \citet{finkelstein2016effect} use a randomized opportunity to apply for Medicaid as an instrument for Medicaid enrollment. They find that half a year of
Medicaid enrollment increases the probability of an emergency department visit by 0.088 percentage points for the compliers. However, the short-term impact of Medicaid on health care utilization may be small, and hence policymakers make efforts to reduce Medicaid coverage disruptions \citep{sugar2021medicaid}. Indeed, our estimated LAFTE shows that continuous Medicaid enrollment for two years increases the probability of an emergency department visit between 0.081 and 0.224 percentage points for the compliers. This suggests that policies encouraging continuous enrollment result in larger effects on health care utilization. We discuss similar policy implications in the other applications. 

The applications also underline the empirical applicability of our methods. The four studies all fit our IV framework, with data available on both enrollment in the first and second part of the treatment. By testing the necessary conditions, we find evidence for the presence of movers in all four studies. In the health program, the movers even establish the majority of the compliers, consistent with the high prevalence of Medicaid coverage disruptions. Except for the development program, the necessary conditions for the double exclusion restriction cannot be rejected. In these three applications, the lower bounds on the LAFTEs are statistically significantly different from zero, and two of the three upper bounds are informative on the size of the LAFTE.  

There is an extensive literature that bounds LATEs under violations of the exclusion restriction. For instance, \cite{flores2013partial} establishes partial identification using one treatment variable. Instead of our approach of using treatment indicators for the first and second part of the treatment, \cite{mealli2013using} use two outcome variables to construct bounds. \cite{conley2012plausibly} propose Bayesian approaches for identification under relaxed exclusion restrictions. \cite{swanson2018partial} provide an overview on treatment bounds for settings with a binary instrument, binary treatment, and a binary outcome. \citet{heckman2000substitution} bound average treatment effects in the presence of control group substitution and treatment group dropping out, without employing instrumental variables.

The identification of causal parameters with multiple treatment parts has been studied in similar settings. First, we study the identification of the LAFTE by indexing the potential outcomes with the full treatment history. Similar potential outcome models are considered in the dynamic treatment literature. For instance, \cite{lechner2009sequential} identifies treatment effects using conditional independence assumptions and \cite{ding2010estimating} employ a structural economic model. \citet{blackwell2017instrumental} identifies the treatment effects for different types of movers between two sequential treatments with one instrument for each treatment. \citet{heckman2016dynamic} also identify dynamic treatment effects using multiple instruments. In contrast, our identification approach relies on only one instrument and does not require a conditionally exogenous treatment or an underlying structural model.

Second, instead of estimating LATEs, recent difference-in-differences techniques are used to estimate the intention to treat in the presence of movers. For instance, \cite{hull2018estimating} identifies mover average treatment effects for the individuals that move in or out of treatment. \cite{verdier2020average} extrapolates treatment effects for movers to stayers: individuals whose treatment status does not change. Specific types of movers are studied in, for example, \citet{athey2022design} who only allow for late-adopters in a staggered adoption framework. 

Third, instead of estimating the LAFTE, causal mediation analysis identifies the direct and indirect effects of enrollment in the first part of treatment. Our proposed double exclusion restriction imposes that all effects of the instrument on the outcome variable go only through enrollment in the first part of treatment. However, the effect of enrollment in the first part on the outcome can either be direct or via treatment enrollment in the second part of the treatment. Hence, enrollment in the second part can be considered a mediator, and \cite{huber2019review} provides an overview of estimation methods for direct and indirect effects of enrollment in the first part of treatment.

The outline of this paper is as follows. \Cref{sec:treatment} defines the causal parameter of interest. \Cref{sec:movers} discusses our IV framework with movers. \Cref{sec:doubleexclusion} introduces the double exclusion restriction, partial identification of the LAFTE, and testable necessary conditions for the presence of movers and the double exclusion restriction. \Cref{sec:application} applies the proposed methods to four empirical treatment evaluation settings. \Cref{sec:conclusion} concludes.

\section{Full treatment effect and instrumental variables}\label{sec:treatment}

Assume a setting in which individuals are randomly assigned to treatment, after which full treatment or only a subset of treatment can be received. The binary instrumental variable $Z \in \{0,1\}$ equals one if an individual is randomly assigned to treatment. The treatment indicator $D_t \in \{0,1\}$ equals one if treatment part $t$ is received, where $t=1$ corresponds to, for instance, college enrollment and $t=2$ to college graduation.
When compliance with the treatment assignment is not mandatory, the treatment indicators may take different values than the instrument. After assignment, individuals may obtain full treatment $\{D_1=1,D_2=1\}$ or no treatment $\{D_1=0,D_2=0\}$. Moreover, $D_1$ may be different from $D_2$ in settings that allow for late enrollment into treatment $\{D_1=0,D_2=1\}$ and/or dropping out of treatment $\{D_1=1,D_2=0\}$. Finally, let the variable $Y$ be the observed outcome of interest. 

\subsection{Causal parameters of interest}\label{sec:causal}

A causal parameter of interest to policy makers is the average treatment effect (ATE) of receiving full treatment $\mathbb{E}[Y(1,1)-Y(0,0)]$, where $Y(D_1,D_2)$ is an individual's potential outcome for values of $D_1$ and $D_2$. This ATE is the average difference between receiving both parts of the treatment ($Y(1,1)$) compared to no treatment at all ($Y(0,0)$). The ATE of the first part of treatment $\mathbb{E}[Y(1,d_2)-Y(0,d_2)]$ or the second part $\mathbb{E}[Y(d_1,1)-Y(d_1,0)]$ may be of interest for the evaluation of which treatment period is most effective. However, most treatments are designed to be received in full.
For instance, the first part of a training program builds knowledge that prepares for the second part, and the second part builds upon the knowledge obtained in the first. Hence, the average treatment effect of receiving full treatment is often of primary interest to policy makers. 

Since we only observe one $Y$ for each individual and individuals may not comply with their assigned treatment, ATEs are in general not identified. In this case it is common to report local ATEs (LATEs) that can be identified by using the random treatment assignment as an instrumental variable for treatment enrollment. The LATE equals the average treatment effect for the individuals who are induced into treatment by the instrument, and is therefore the ATE for a subpopulation.

\subsection{The standard IV framework}

The standard IV framework introduced by \cite{imbens1994identification} defines an IV estimand and shows that it identifies a LATE of a treatment variable $D$ on the outcome of interest $Y$ using the potential outcome framework and four assumptions on the instrumental variable $Z$. The IV estimand is defined as 
\begin{align}\label{eq:IV}
    \beta_{IV}(D) =\frac{\Delta\mathbb{E}[Y|Z]}{\Delta\mathbb{E}[D|Z]},
\end{align}
where $\Delta \mathbb{E}[A|Z]=\mathbb{E}[A|Z=1]-\mathbb{E}[A|Z=0]$. The potential outcome framework links the observed treatment indicator $D$ to the potential treatment status $D(Z)$ as $D=D(1)Z+D(0)(1-Z)$, and the observed outcome $Y$ to the potential outcomes $Y(Z,D)$ as $Y=Y(1,1)ZD+Y(1,0)Z(1-D)+Y(0,1)(1-Z)D+Y(0,0)(1-Z)(1-D)$. 

The four instrumental variable assumptions are:
\begin{assumption}[Instrumental variable assumptions]\label{ass:iv}\
\begin{enumerate}
    \item (Independence) $Y(z,d), D(z) \perp Z$ for all $z,d$.
    \item (Relevance) $\Delta\mathbb{E}[D|Z] > 0$.
    \item (Monotonicity) $D(1)\geq D(0)$.
    \item (Exclusion) $Y(z,d)=Y(d)$ for all $z,d$. 
\end{enumerate} 
\end{assumption}
Under \Cref{ass:iv}, the IV estimand in \eqref{eq:IV} identifies a LATE if the treatment can be summarized by a single binary variable $D$:  $\beta_{IV} = \mathbb{E}[Y(1)-Y(0)|C]$, where $C=\{D(1)-D(0)=1\}$ defines the individuals induced into treatment by the instrument. These individuals are referred to as compliers.

\Cref{tab:treatment_def} shows four ways the treatment indicator $D$ can be constructed when $\{Z,D_1,D_2,Y\}$ is observed. The treatment indicator can be defined as enrollment at the start of the treatment ($D_1$), enrollment at the end of the treatment ($D_2$), received full treatment ($D_\land$), or received at least one part of the treatment ($D_\lor$). The third column in \Cref{tab:treatment_def} shows that different definitions of the treatment indicator are used in the applied economics literature. 

\begin{table}[t]
\small
\begin{center}
\caption{Treatment indicator specifications}
\label{tab:treatment_def}
\begin{tabular}{ l l p{7.3cm}} 
 \toprule \toprule
 Definition $D$ & Treatment indicator & Examples in empirical work \\
 & for enrollment in ... &  \\
 \midrule
 $D_1$ & first part  &  \cite{silliman2022labor}, \cite{grosz2020returns}, \cite{zimmerman2014returns}, \cite{burde2013bringing}, \cite{hoekstra2009effect} \\
 $D_2$ & second part  &  \cite{ketel2016returns}, \cite{zimmerman2014returns} \\
 $D_\land=D_1D_2$ & both parts  & \\ 
 $D_\lor=D_1+D_2-D_1D_2$ & at least one part &    \cite{wheeler2022linkedin}, \cite{anelli2020returns}, \cite{finkelstein2012oregon}\\
 \bottomrule\bottomrule
\end{tabular}
        \begin{tablenotes}
	\footnotesize
	\item[] Notes: this table provides an overview of four treatment indicators that can be constructed when $\{D_1,D_2\}$ is observed and gives examples of research articles using each treatment indicator. 
        \end{tablenotes}
\end{center}
\end{table}

\section{Instrumental variables with movers}\label{sec:movers}

When the instrument induces individuals to obtain only a subset of the treatment, all four definitions for $D$ may result in violations of the exclusion restriction made by \Cref{ass:iv}.4. These violations arise when $Z$ affects $Y$ while $D$ remains constant. For instance, if the researcher uses $D=D_1$ and the instrument induces some individuals into only the second part of treatment, $Z$ may affect $Y$ through $D_2$ whereas $D_1$ is not affected. Similarly, if the researcher uses $D=D_2$, the exclusion restriction may be violated if the instrument induces some individuals into only the first part of treatment. An instrument inducing individuals from no treatment at all to only part of the treatment, or an instrument inducing individuals from only part of the treatment to full treatment, may run into problems with $D=D_\land$ or $D=D_\lor$ respectively.

This section extends the standard IV framework to include the potential treatment status at the first and second part of the treatment. Within this framework, five different groups of compliers can be distinguished. The four additional groups of compliers cause the violations of the exclusion restriction described above. We show that in the presence of these four groups, LATEs can only be point identified under additional strict assumptions. 

\subsection{Model and notation} 

We extend the potential outcome model to include two potential treatments $D_1(Z)$ and $D_2(Z,D_1)$ and a potential outcome $Y(Z,D_1,D_2)$. Potential and observed treatments and outcomes are linked as follows,
\begin{align}
\label{eq:po_d1}
 D_1 =& D_1(1)Z+D_1(0)(1-Z),\\
 \label{eq:po_d2}
 D_2 =& [D_2(1,1)D_1+D_2(1,0)(1-D_1)]Z + [D_2(0,1)D_1+D_2(0,0)(1-D_1)](1-Z), \\
 \label{eq:po_y}
 Y   =& [Y(1,1,1)D_1D_2+Y(1,0,1)(1-D_1)D_2+Y(1,1,0)D_1(1-D_2)+ \\&\,\,Y(1,0,0)(1-D_1)(1-D_2)]Z + \notag[Y(0,1,1)D_1D_2+Y(0,0,1)(1-D_1)D_2+\\&\,\,Y(0,1,0)D_1(1-D_2)+Y(0,0,0)(1-D_1)(1-D_2)](1-Z). \notag
\end{align}
\Cref{ass:iv2} replaces \Cref{ass:iv} for the potential outcome framework defined above:
\begin{assumption}[Instrumental variable assumptions]\label{ass:iv2}\
\begin{enumerate}
    \item (Independence) $Y(z,d_1,d_2), D_1(z), D_2(z,d_1) \perp Z$ for all $z,d_1,d_2$.
    \item (Relevance) $\Delta\mathbb{E}[D|Z] > 0$ for $D=D_1,D_2,D_{\land},D_{\lor}$.
    \item (Monotonicity) $D_1(1)\geq D_1(0)$ and $D_2(1,D_1(1))\geq D_2(0,D_1(0))$.
    \item (Exclusion) $Y(z,d_1,d_2)=Y(d_1,d_2)$ for all $z,d_1,d_2$. 
\end{enumerate} 
\end{assumption}

These assumptions hold if $Z$ is randomly assigned (\ref{ass:iv2}.1), if there are individuals that are induced into full treatment by the instrument (\ref{ass:iv2}.2), and there are no individuals induced to move out of any part of treatment by the instrument (\ref{ass:iv2}.3). \Cref{ass:iv2}.4 implies that 
\begin{align}\label{eq:Yd}
Y   = Y(1,1)D_1D_2+Y(0,1)(1-D_1)D_2+Y(1,0)D_1(1-D_2)+Y(0,0)(1-D_1)(1-D_2),
\end{align}
which shows that $Y$ is a function of the two treatment parts, as defined in \Cref{sec:causal}. 

In contrast, the exclusion restriction in \Cref{ass:iv}.4 imposes that $Y(z,d_1,d_2)=Y(d)$ for all $z,d_1,d_2$, which only depends on one summary of the treatment $D=d$. \Cref{ass:iv}.4 is more restrictive as it does not allow $Z$ to affect $Y$ through both treatment parts, whereas \Cref{ass:iv2}.4 does. For instance, with $D=D_1$, \Cref{ass:iv}.4 imposes that $Y(z,d_1,d_2)=Y(d_1)$ and therefore $Y$ does not depend on the second treatment part. 

The potential outcome model in \eqref{eq:Yd} implies that there are different types of compliers. Since individuals can comply with the instrument in the first part, the second part, or the full treatment, the group of compliers $C$ consists of five different complier groups: $C=\{\{C_1,C_2\},\{C_1,N_2\},\{C_1,A_2\},\{N_1,C_2\},\{A_1,C_2\}\}$. These groups are defined as follows:
\begin{enumerate}
    \item $\{C_1,C_2\} = \{D_1(1)-D_1(0)=1,D_2(1,1)-D_2(0,0)=1\},$
    \item $\{C_1,N_2\} = \{D_1(1)-D_1(0)=1,D_2(1,1)=D_2(0,0)=0\},$
    \item $\{C_1,A_2\} = \{D_1(1)-D_1(0)=1,D_2(1,1)=D_2(0,0)=1\},$
    \item $\{N_1,C_2\} = \{D_1(1)=D_1(0)=0,D_2(1,0)-D_2(0,0)=1\},$
    \item $\{A_1,C_2\} = \{D_1(1)=D_1(0)=1,D_2(1,1)-D_2(0,1)=1\},$
\end{enumerate}
where $C_t$, $N_t$, and $A_t$ refer to respectively compliers, never-takers, and always-takers in treatment part $t$. The first group of compliers $\{C_1,C_2\}$ is induced into full treatment by the instrument, and we refer to this group as \textit{full compliers}. We define the individuals that are induced to obtain only part of the treatment by the instrument as \textit{movers}. Hence, there are four types of movers: $\{C_1,N_2\}$ and $\{C_1,A_2\}$ are induced into only the first part of treatment, whereas $\{N_1,C_2\}$ and $\{A_1,C_2\}$ are induced into only the second part of treatment.

Our definition of movers concerns individuals who change treatment status after the start of the program depending on the value of the instrument. The two mover types $\{C_1,N_2\}$ and $\{A_1,C_2\}$ dropout from treatment and miss the second part of the program when $Z=1$ and $Z=0$, respectively. The two mover types $\{C_1,A_2\}$ and $\{N_1,C_2\}$ are late-adopters and miss the first part of the program when $Z=0$ and $Z=1$, respectively. Individuals with $\{N_1,A_2\}$ or $\{A_1,N_2\}$ are not included, as they do not affect the LATE identification. Hence, our mover group is different from the more general definition of movers as the set of all individuals with $D_1 \neq D_2$, as used by, for instance, \citet{hull2018estimating} in difference-in-differences estimation. 

Using the potential outcome framework in \eqref{eq:Yd} and the definition of the complier group $C$, the local average full treatment effect (LAFTE) can now be expressed as $\mathbb{E}[Y(1,1)-Y(0,0)|C]$.

\subsection{What IV can identify with movers}

The following lemma shows what the first stages $\Delta\mathbb{E}[D|Z]$ for different definitions of $D$ and the reduced form $\Delta\mathbb{E}[Y|Z]$ identify when the potential outcome framework takes both treatment parts into account.
\begin{lemma}[First stages and reduced form]\label{lemma:fs_rf}\ \\
Under \Cref{ass:iv2} it holds that
\begin{enumerate}
    \item $\Delta\mathbb{E}[D_1|Z]=\mathbb{P}[C_1,C_2]+\mathbb{P}[C_1,N_2]+\mathbb{P}[C_1,A_2]$.
    \item $\Delta\mathbb{E}[D_2|Z]=\mathbb{P}[C_1,C_2]+\mathbb{P}[N_1,C_2]+\mathbb{P}[A_1,C_2]$.
    \item $\Delta\mathbb{E}[D_\land|Z]=\mathbb{P}[C_1,C_2]+\mathbb{P}[C_1,A_2]+\mathbb{P}[A_1,C_2]$.
    \item $\Delta\mathbb{E}[D_\lor|Z]=\mathbb{P}[C_1,C_2]+\mathbb{P}[C_1,N_2]+\mathbb{P}[N_1,C_2]$.
    \item $\begin{aligned}[t]
    \Delta\mathbb{E}[Y|Z] =&\, \mathbb{E}[Y(1,1)-Y(0,0)|C_1,C_2]\mathbb{P}[C_1,C_2]+
    \mathbb{E}[Y(1,0)-Y(0,0)|C_1,N_2]\mathbb{P}[C_1,N_2]+ \\
    &\, \mathbb{E}[Y(1,1)-Y(0,1)|C_1,A_2]\mathbb{P}[C_1,A_2]+
    \mathbb{E}[Y(0,1)-Y(0,0)|N_1,C_2]\mathbb{P}[N_1,C_2]+ \\
    &\, \mathbb{E}[Y(1,1)-Y(1,0)|A_1,C_2]\mathbb{P}[A_1,C_2].
\end{aligned}$
\end{enumerate}
\end{lemma}
The proof is deferred to \Cref{A:proof_lem1}.

\Cref{lemma:fs_rf} shows that the first stages represent the proportion of different complier types depending on the definition of $D$. Each first stage captures the proportion of the full compliers plus the proportions of two mover types that comply with the corresponding $D$. For instance, with $D=D_1$, the movers $\{C_1,N_2\}$ and $\{C_1,A_2\}$ comply in the first part of treatment. 

The reduced form coefficient $\Delta\mathbb{E}[Y|Z]$ equals a weighted sum of LATEs, in which each weight corresponds to the proportion of one of the five complier types. Each LATE corresponds to a different treatment effect for a complier type. The first term of the reduced form in \Cref{lemma:fs_rf} corresponds to the LAFTE, but only for the full compliers $\{C_1,C_2\}$. 

For each of the definitions of $D$ in \Cref{tab:treatment_def}, the IV estimand in \eqref{eq:IV} identifies a weighted average of LATEs plus a bias term. For instance, for $D=D_1$ we have
\begin{align}
\left.\begin{aligned}
 \beta_{IV}(D_1) =  &\mathbb{E}[Y(1,1)-Y(0,0)|C_1,C_2]w(C_1,C_2)+\\
        &\mathbb{E}[Y(1,0)-Y(0,0)|C_1,N_2]w(C_1,N_2)+\\
    &\mathbb{E}[Y(1,1)-Y(0,1)|C_1,A_2]w(C_1,A_2)+\end{aligned}\right\}       \text{ weighted LATEs} \\
    \left.\begin{aligned}
 &\mathbb{E}[Y(1,1)-Y(1,0)|N_1,C_2]w(N_1,C_2)+ \notag \\
 &\mathbb{E}[Y(0,1)-Y(0,0)|A_1,C_2]w(A_1,C_2),\end{aligned}\right\}       \text{bias \hspace{0.82in}} 
\end{align}
with $w(G_1,G_2)=\frac{\mathbb{P}[G_1,G_2]}{\mathbb{P}[C_1,C_2]+\mathbb{P}[C_1,N_2]+\mathbb{P}[C_1,A_2]}$, and where $\{G_1,G_2\}$ can represent any of the five complier groups. The bias terms reflect the effect of $D_2$ on $Y$ while keeping $D_1$ constant. It follows that the IV estimand does not have a clear causal interpretation:

\begin{proposition}[No causal interpretation of $\beta_{IV}(D)$]\label{theorem:standardIV}\ \\
Under \Cref{ass:iv2} it holds that the IV estimand in \eqref{eq:IV} has no causal interpretation with either $D=D_1,D_2,D_\land,D_\lor$ as defined in \Cref{tab:treatment_def}. 
\end{proposition}
The proof follows directly from \Cref{lemma:fs_rf}.

Figure~\ref{fig:switchers} visualizes the identification problem, where each arrow represents a possible effect among $\{Z,D_1,D_2,Y\}$ under \Cref{ass:iv2}. The instrument induces the full compliers $\{C_1,C_2\}$ into full treatment, but also induces the mover groups into only a subset of the treatment. The figure suggests that the LAFTE can be identified under two different additional assumptions. The first assumption rules out  the presence of all mover groups. The second assumes that the treatment effects for movers are identical to their full treatment effect. \Cref{cor:no_movers} and \ref{cor:homo_movers} below formalize this intuition.

\begin{figure}[t]
\centering
\caption{An overview of the possible effects in IV with movers.}
\label{fig:switchers}
\begin{tikzpicture}[shorten >=2pt,node distance=7.5cm,on grid,auto]
   \node[] (J) {$Z$};   
   \node[right=50mm of J] (K) {$D_1(z)$};   
   \node[right=50mm of K] (L) {$D_2(z,d_1)$};
   \node[right=50mm of L] (M)  {$Y(d_1,d_2)$};
   \path[->]
    (J) edge [left]  node {} (K) 
    (K) edge [left]  node {} (L)
    (L) edge [left]  node [above] {$\{C_1,C_2\}$ } (M)
    (J) edge [bend left, dashed]  node[above] {Double exclusion} node[below] {$\{N_1,C_2\},\{A_1,C_2\}$} (L)
    (K) edge [bend right] node {$\{C_1,N_2\},\{C_1,A_2\}$} (M);   
\end{tikzpicture}
\caption*{ \footnotesize
Notes: under \Cref{ass:iv2}, $Z$ cannot be affected by any variable, $D_1(z)$ can only be affected by $Z$, $D_2(z,d_1)$ can only be affected by both $Z$ and $D_1$, and $Y(d_1,d_2)$ can only be affected by both $D_1$ and $D_2$. The dashed arrow represents the effect prevented by \Cref{ass:double_exclusion}.}
\end{figure}

\begin{corollary}[No movers]\label{cor:no_movers}\ \\
Under \Cref{ass:iv2} and $\mathbb{P}[C_1,N_2]=\mathbb{P}[C_1,A_2]=\mathbb{P}[N_1,C_2]=\mathbb{P}[A_1,C_2]=0$, it holds that
\begin{align}
    \beta_{IV}(D)=\frac{\Delta\mathbb{E}[Y|Z]}{\Delta\mathbb{E}[D|Z]}=\mathbb{E}[Y(1,1)-Y(0,0)|C],
\end{align}
for $D=D_1,D_2,D_\land,D_\lor$.
\end{corollary}
The proof follows directly from \Cref{lemma:fs_rf}. \Cref{cor:no_movers} may suit a setting in which, for instance, treatment completion is mandatory after treatment enrollment and treatment completion is impossible without treatment enrollment. It follows that all compliers are full compliers $C=\{C_1,C_2\}$.

\begin{corollary}[Homogeneous treatment effects movers]\label{cor:homo_movers}\ \\
Under \Cref{ass:iv2} and the treatment effect homogeneity assumptions in \Cref{A:proof_cor2}, it holds that the IV estimand in \eqref{eq:IV} with $D=D_1,D_2,D_\land,D_\lor$ identifies the LAFTE for three complier groups.
\end{corollary}
The proof is deferred to \Cref{A:proof_cor2}. \Cref{cor:homo_movers} may suit a setting in which, for instance, $\mathbb{E}[Y(1,d_2)-Y(0,d_2)|G_1,G_2]$ does not depend on $d_2$ for the movers $\{G_1,G_2\}$. This setting implies homogeneous treatment effects within mover types. 

The LAFTE can also be identified by a combination of assumptions on the presence of certain mover types and the homogeneity of certain treatment effects. This is formalized by the dynamic exclusion model discussed by \citet{angrist2022marginal}:
\begin{align}\label{eq:de_d1}
   D_1=&\delta+\pi Z + \eta,\\
   D_2=&\alpha+\psi D_1 +\xi, \label{eq:de_d2}\\
    Y =&\beta+\mu D_2 + \varepsilon,  \label{eq:de_y}
\end{align}
where $\delta$, $\pi$, $\alpha$, $\psi$, $\beta$, and $\mu$ are coefficients, $\eta$, $\xi$, and $\varepsilon$ are error terms, and we exclude additional covariates. From \eqref{eq:de_d2} follows that $D_2$ does not depend on $Z$, and consequently the movers $\{N_1,C_2\}$ and $\{A_1,C_2\}$ in \Cref{fig:switchers} do not exist. Similarly, \eqref{eq:de_y} shows that $Y$ does not depend on $D_1$. \Cref{fig:switchers} shows that this is the case if either $\{C_1,N_2\}$ and $\{C_1,A_2\}$ do not exist, or if $Y(0,d_2)=Y(1,d_2)$ for all $d_2$.

\section{Movers with a double exclusion restriction}\label{sec:doubleexclusion}

This section proposes a partial identification strategy for the LAFTE. We replace the single exclusion restriction in \Cref{ass:iv2}.4, by a double exclusion restriction:
\begin{assumption}[Double exclusion]\label{ass:double_exclusion}\
\begin{enumerate}
\item[] $Y(z,d_1,d_2)=Y(d_1,d_2)$ for all $z,d_1,d_2$ and $D_2(z,d_1)=D_2(d_1)$ for all $z,d_1$. 
\end{enumerate}
\end{assumption}

The first part of \Cref{ass:double_exclusion} is identical to \Cref{ass:iv2}.4, and the second part states that $Z$ must not have a direct effect on $D_2$ other than through $D_1$. These two parts together impose a double exclusion restriction on $Z$. 

Figure~\ref{fig:switchers} shows that the double exclusion allows for $Z$ to have a direct effect on $D_1$, and for $Z$ to have an effect on $D_2$ through $D_1$. Hence, compliers in the second part of treatment have to be compliers in the first part, and the group of compliers now consists of only three types: $C=\{\{C_1,A_2\},\{C_1,N_2\},\{C_1,C_2\}\}$. \Cref{ass:double_exclusion} holds under \eqref{eq:de_d2}, but does not impose \eqref{eq:de_y}, and hence does not exclude all mover types or imposes homogeneous treatment effects.

The double exclusion restriction imposes that $Z$ randomly assigns individuals to the first part of treatment but not the second part of treatment if the first part stays constant. This holds in settings in which a delayed treatment response to treatment assignment can be ruled out. For instance, in the Medicaid experiment analysed by \cite{finkelstein2012oregon}, individuals with a lottery draw of $Z=1$ could only obtain Medicaid at the start of the treatment, and hence mover type $\{N_1,C_2\}$ is likely to be absent. Since individuals with $Z=0$ had little to no opportunity to obtain medicaid coverage at the start of the treatment, mover type $\{A_1,C_2\}$ is also likely to be absent. In case the first stage estimate for enrollment at the start of treatment is close to one, for instance in a carefully conducted randomized controlled trial \citep{duflo2007using,deree2023closing}, the double exclusion may also hold. 

\subsection{Partial identification of the LAFTE}

Since the double exclusion restriction rules out two complier types, $\{A_1,C_2\}$ and $\{N_1,C_2\}$, the first stages and reduced form in \Cref{lemma:fs_rf} can be simplified. 

\begin{lemma}[First stages and reduced form with double exclusion]\label{lemma:fs_rf_double}\ \\
Under \Cref{ass:iv2}.1-\ref{ass:iv2}.3 and \ref{ass:double_exclusion} it holds that
\begin{enumerate}
    \item $ \Delta\mathbb{E}[D_1|Z]=\mathbb{P}[C_1,C_2]+\mathbb{P}[C_1,N_2]+\mathbb{P}[C_1,A_2]$.
    \item $ \Delta\mathbb{E}[D_2|Z]=\mathbb{P}[C_1,C_2]$.
    \item $ \Delta\mathbb{E}[D_\land|Z]=\mathbb{P}[C_1,C_2]+\mathbb{P}[C_1,A_2]$.
    \item $\Delta\mathbb{E}[D_\lor|Z]=\mathbb{P}[C_1,C_2]+\mathbb{P}[C_1,N_2]$.
    \item $\begin{aligned}[t]
    \Delta\mathbb{E}[Y|Z] =&\, \mathbb{E}[Y(1,1)-Y(0,0)|C_1,C_2]\mathbb{P}[C_1,C_2]+
     \mathbb{E}[Y(1,0)-Y(0,0)|C_1,N_2]\mathbb{P}[C_1,N_2]+\\
     &\,\mathbb{E}[Y(1,1)-Y(0,1)|C_1,A_2]\mathbb{P}[C_1,A_2].
\end{aligned}$
\end{enumerate}
\end{lemma}
The proof is deferred to \Cref{A:proof_lem3}. \Cref{lemma:fs_rf_double} shows that under the double exclusion restriction the proportions of all three complier groups are identified. 

The identification of the proportions of the complier groups allows for the partial identification of LAFTE under additional assumptions:

\begin{theorem}[Partial identification LAFTE]\label{theorem:late_c}\ \\
Under \Cref{ass:iv2}.1-\ref{ass:iv2}.3 and \ref{ass:double_exclusion}, positive response $\mathbb{E}[Y(0,0)|C_1,A_2]\geq 0$, monotone treatment responses $\mathbb{E}[Y(1,1)-Y(1,0)|C_1,N_2]\geq 0 \text{ and } \mathbb{E}[Y(0,1)-Y(0,0)|C_1,A_2]\geq 0,$ and monotone treatment selections $\mathbb{E}[Y(1,1)|C_1,A_2] \geq \mathbb{E}[Y(1,1)|C_1,N_2] \text{ and } \mathbb{E}[Y(1,1)|C_1,C_2] \geq \mathbb{E}[Y(1,1)|C_1,N_2]$, it holds that
\begin{align}
        \frac{\Delta\mathbb{E}[Y|Z]}{\Delta\mathbb{E}[D_1|Z]}\leq  \mathbb{E}[Y(1,1)-Y(0,0)|C]  \leq 
  \frac{\Delta\mathbb{E}[D_\land Y|Z]}{\Delta\mathbb{E}[D_\land|Z]}+\frac{\Delta\mathbb{E}[(1-D_1)(1-D_2)Y|Z]}{\Delta\mathbb{E}[D_1|Z]}.
\end{align}
\end{theorem}
The proof is deferred to \Cref{A:proof_theorem3}. Since the bounds in \Cref{theorem:late_c} collapse to the LAFTE for the full compliers in absence of movers, the bounds are sharp.

Assuming that the LATEs $\mathbb{E}[Y(1,1)-Y(1,0)|C_1,N_2]$ and $\mathbb{E}[Y(0,1)-Y(0,0)|C_1,A_2]$ are nonnegative, the difference between the reduced form of the LAFTE and the reduced form in \Cref{lemma:fs_rf_double} is positive. Hence, the latter can be used as a lower bound on the LAFTE. The assumption of positive expected treatment effects have been used by, for instance, \citet{flores2013partial} for the partial identification of LATEs with a single treatment indicator. \citet{manski1997monotone} introduces the monotone treatment response (MTR) assumption on the individual level instead of in expectation. 

The reduced form in \Cref{lemma:fs_rf_double} only differs from the reduced form of the LAFTE by the two potential outcomes $\mathbb{E}[Y(0,0)|C_1,A_2]$ and $\mathbb{E}[Y(1,1)|C_1,N_2]$. The upper bound on the LAFTE can be obtained by bounding these potential outcomes using the following assumptions: The potential outcome $\mathbb{E}[Y(0,0)|C_1,A_2]$ is non-negative, and the potential outcome of obtaining full treatment $Y(1,1)$ is smaller for $\{C_1,N_2\}$ than for $\{C_1,C_2\}$ and $\{C_1,A_2\}$. Outcomes can generally be rescaled so that the assumption of a non-negative potential outcome is harmless, for instance when the response has a lower bound. The second assumption is invoked in expectation, and therefore weaker than the monotone treatment selection (MTS) assumption of \citet{manski2000monotone}. 

A combination of assumptions similar to the ones in \Cref{theorem:late_c} have been used for the partial identification of treatment effects by, for instance, \citet{molinari2010missing}, \citet{dehaan2011effect}, and \citet{kreider2012identifying}. In general, partial identification is common in the analysis of treatment effect identification problems. For instance, \citet{kreider2007disability}, \citet{battistin2011misclassified}, \citet{tommasi2020bounding}, and \citet{calvi2022late} derive bounds on treatment effects when the treatment is misreported.

When the MTR and MTS assumptions are considered too strong, the following bounds can be derived assuming only a bounded response:
\begin{corollary}[Partial identification LAFTE with bounded response]\label{corr:bounded}\ \\
Under \Cref{ass:iv2}.1-\ref{ass:iv2}.3 and \ref{ass:double_exclusion}, with bounded response: $Y \in [Y_{\min}, Y_{\max}]$, it holds that
    \begin{align}
    & \frac{\Delta\mathbb{E}[(1-D_1-D_2+2D_1D_2)Y|Z]+Y_{\min}\Delta\mathbb{E}[D_\lor-D_2|Z]-Y_{\max}\Delta\mathbb{E}[D_\land-D_2|Z]}{\Delta\mathbb{E}[D_1|Z]}\\  \notag
&\leq\mathbb{E}[Y(1,1)-Y(0,0)|C]\leq\\  \notag
   & \frac{\Delta\mathbb{E}[(1-D_1-D_2+2D_1D_2)Y|Z]+Y_{\max}\Delta\mathbb{E}[D_\lor-D_2|Z]-Y_{\min}\Delta\mathbb{E}[D_\land-D_2|Z]}{\Delta\mathbb{E}[D_1|Z]}.
\end{align}
\end{corollary}
The proof is deferred to \Cref{A:corr_bounded}. 

The bounded response allows us to replace $Y(1,0)$ for $\{C_1,N_2\}$ and $Y(0,1)$ for $\{C_1,A_2\}$ with respectively  $Y_{\min}$ and $Y_{\max}$ ($Y_{\max}$ and $Y_{\min}$) to construct a lower (upper) bound on the LAFTE. The difference between the bounds equals 
$(y_{\max}-y_{\min})(\mathbb{P}[C_1,N_2]+\mathbb{P}[C_1,A_2])/(\mathbb{P}[C_1,C_2]+\mathbb{P}[C_1,N_2]+\mathbb{P}[C_1,A_2])$. The bounds equal $\mathbb{E}[Y(1,1)-Y(0,0)|C_1,C_2]$ when no movers are present, but can be wide when the response bounds are wide and the proportion of movers is large. 

A complier group that may be of particular interest are the full compliers $\{C_1,C_2\}$. This group is induced to obtain both parts of the treatment by the instrument. However, with only one instrument, it is not possible to affect the full compliers without also affecting the other complier types. Hence, the full compliers are mostly relevant for settings in which the policymaker has an instrument for both the first and second part of treatment. \cite{blackwell2017instrumental} shows that with two binary instruments, and an additional treatment exclusion restriction that each instrument only affects its own treatment, the LAFTE for the full compliers can be identified.

\subsection{Testing assumptions}\label{sec:testing}

Our identification strategy for the LAFTE relies on restrictions on the presence of mover types. This section derives testable necessary conditions for these restrictions.

First, \Cref{cor:no_movers} shows that the IV method introduced by \cite{imbens1994identification} point-identifies the LAFTE when no movers are present. This assumption results in the following necessary conditions.
\begin{proposition}[Necessary conditions no movers]\label{prop:nec_cond_movers}\ \\
Under \Cref{ass:iv2} it holds that 
\begin{align}
    \Delta\mathbb{E}[D_{\lor}-D_2|Z] &=\mathbb{P}[C_1,N_2]-\mathbb{P}[A_1,C_2],\label{eq:nC_1}\\
    \Delta\mathbb{E}[D_{\land}-D_2|Z] &=\mathbb{P}[C_1,A_2]-\mathbb{P}[N_1,C_2],\label{eq:nC_2}
\end{align}
which both equal zero if \, $\mathbb{P}[C_1,N_2]=\mathbb{P}[A_1,C_2]$ and $\mathbb{P}[C_1,A_2]=\mathbb{P}[N_1,C_2]$.

Under \Cref{ass:iv2} and $p_1=\mathbb{P}[C_1,N_2]=\mathbb{P}[A_1,C_2]$ and $p_2=\mathbb{P}[C_1,A_2]=\mathbb{P}[N_1,C_2]$, it holds that
\begin{align}
     \Delta\mathbb{E}[(D_{\lor}-D_2)Y|Z] &= (\mathbb{E}[Y(1,0)|C_1,N_2]-\mathbb{E}[Y(1,0)|A_1,C_2])p_1,\label{eq:nc3}\\
    \Delta\mathbb{E}[(D_{\land}-D_2)Y|Z] &= (\mathbb{E}[Y(0,1)|C_1,A_2]-\mathbb{E}[Y(0,1)|N_1,C_2])p_2,\label{eq:nc4}
\end{align}
which both equal zero if \, $\mathbb{P}[C_1,N_2]=\mathbb{P}[A_1,C_2]=\mathbb{P}[C_1,A_2]=\mathbb{P}[N_1,C_2]=0$ and/or\\ $\mathbb{E}[Y(1,0)|C_1,N_2]=\mathbb{E}[Y(1,0)|A_1,C_2]$ and $\mathbb{E}[Y(0,1)|C_1,A_2]-\mathbb{E}[Y(0,1)|N_1,C_2]$.
\end{proposition}
The proof is deferred to \Cref{A:proof_prop1}.

The first two necessary conditions \eqref{eq:nC_1} and \eqref{eq:nC_2} in \Cref{prop:nec_cond_movers} can be used to test the null-hypothesis that $\mathbb{P}[C_1,N_2]=\mathbb{P}[A_1,C_2]$ and $\mathbb{P}[N_1,C_2]=\mathbb{P}[C_1,A_2]$. A failure to reject this null-hypothesis does not necessarily imply the absence of movers, as it may also indicate mover types with identical proportions. Hence, in case of a failure to reject \eqref{eq:nC_1} and \eqref{eq:nC_2}, the necessary conditions \eqref{eq:nc3} and \eqref{eq:nc4} can be considered. These conditions are equal to zero if there are no movers or if potential outcomes are homogeneous.

Provided that potential outcomes are heterogeneous, \Cref{prop:nec_cond_movers} provides a two-step testing procedure for the presence of movers. If the null-hypothesis that \eqref{eq:nC_1} and \eqref{eq:nC_2} both equal zero is rejected, movers may be present. If this null-hypothesis cannot be rejected, the null-hypothesis that \eqref{eq:nc3} and \eqref{eq:nc4} both equal zero has to be tested. In case of a rejection, we still conclude that there may be movers. However, a failure to reject indicates that there are no movers. So if both sets of necessary conditions cannot be rejected, we recommend using standard IV approaches to identify the LAFTE for the full compliers $\{C_1,C_2\}$. 

Second, \Cref{theorem:late_c} shows that in the absence of certain mover types, the LAFTE is partially identified. In particular, the double exclusion restriction in \Cref{ass:double_exclusion} only allows for $\{C_1,A_2\}$ and $\{C_1,N_2\}$. This has sign implications for \eqref{eq:nC_1} and \eqref{eq:nC_2}: 

\begin{proposition}[Necessary condition double exclusion restriction]\label{prop:nec_cond_double}\ \\
Under \Cref{ass:iv2} it holds that 
\begin{align}
    \Delta\mathbb{E}[D_{\lor}-D_2|Z] &=\mathbb{P}[C_1,N_2]-\mathbb{P}[A_1,C_2], \label{eq:nc5}\\
    \Delta\mathbb{E}[D_{\land}-D_2|Z] &=\mathbb{P}[C_1,A_2]-\mathbb{P}[N_1,C_2], \label{eq:nc6}
\end{align}
which are both nonnegative if \Cref{ass:double_exclusion} holds.
\end{proposition}
The proof follows directly from \Cref{prop:nec_cond_movers}.

The sign conditions in \Cref{prop:nec_cond_double} are necessary for the absence of the movers ruled out by the double exclusion restriction. Since it could be the case that $\mathbb{P}[C_1,N_2]>\mathbb{P}[A_1,C_2]>0$ and $\mathbb{P}[C_1,A_2]>\mathbb{P}[N_1,C_2]>0$, the conditions are not sufficient for the double exclusion restriction to hold. Hence, a failure to reject the null-hypothesis that the sign restrictions hold need not imply the double exclusion restriction, but a rejection implies that the double exclusion restriction cannot be invoked. 

\subsection{Causal estimands under weaker assumptions}\label{sec:empiricalwork}

\Cref{theorem:late_c} provides sharp nonparametric bounds for the LAFTE. However, point-identification may be required or in some empirical settings the assumptions may be deemed too strong. For these cases, we discuss three alternative causal objects that can be identified under weaker assumptions, and from which two can be point-identified. 

First, \Cref{lemma:fs_rf_double} shows that under \Cref{ass:iv2}.1-\ref{ass:iv2}.3 and \ref{ass:double_exclusion}, the IV estimand in \eqref{eq:IV} with $D=D_1$ identifies the LATE of taking the first part of the treatment:
\begin{align}\label{eq:iv_d1_afterdouble}
    \beta_{IV}(D_1) = &\mathbb{E}[Y(1,1)-Y(1,0)|C_1,C_2]w[C_1,C_2] +\mathbb{E}[Y(1,0)-Y(0,0)|C_1,C_2]w[C_1,C_2]+  \\ \notag
    &\,\mathbb{E}[Y(1,0)-Y(0,0)|C_1,N_2]w[C_1,N_2]+ 
    \mathbb{E}[Y(1,1)-Y(0,1)|C_1,A_2]w[C_1,A_2],
\end{align}
with $w[G_1,G_2]=\frac{\mathbb{P}[G_1,G_2]}{\mathbb{P}[C_1,C_2]+\mathbb{P}[C_1,N_2]+\mathbb{P}[C_1,A_2]}$.

The weighted average of LATEs in \eqref{eq:iv_d1_afterdouble} can be interpreted as a decomposition of the total effect of $D_1$ on $Y$ using $D_2$ as a mediator. The mediation literature, see \cite{huber2019review} for a review, defines the direct effect as the effect of $D_1$ on $Y$ while the mediator $D_2$ is constant, and the indirect effect as the effect of $D_2$ on on $Y$ while $D_1$ is constant. Hence, the first term in \eqref{eq:iv_d1_afterdouble} is the indirect effect, and the remaining terms are direct effects.

\Cref{tab:treatment_def} shows that the IV estimand in \eqref{eq:iv_d1_afterdouble} is often used in the literature. This paper shows that this is a valid estimator of the effect of $D_1$ under the double exclusion restriction. In addition, we argue that the effect of receiving full treatment is also a causal parameter of interest. Adding $\mathbb{E}[Y(1,1)-Y(1,0)|C_1,N_2]w[C_1,N_2]$ and $\mathbb{E}[Y(0,1)-Y(0,0)|C_1,A_2]w[C_1,A_2]$ to \eqref{eq:iv_d1_afterdouble} results in the effect of receiving full treatment. The additional MTR assumptions in \Cref{theorem:late_c} guarantee that these two average treatment effects of receiving the second part of treatment are positive. It follows that the lower bound on the LAFTE equals \eqref{eq:iv_d1_afterdouble} and the LAFTE is equal to or larger than the LATE of taking the first part of the treatment.

Second, the double exclusion restriction may be considered too strong, or its necessary conditions may be rejected for the setting at hand. \Cref{lemma:fs_rf_double} shows that under only \Cref{ass:iv2}, the IV estimand in \eqref{eq:IV} with the multivalued treatment $D=D_1+D_2 \in \{0,1,2\}$ equals
\begin{align}\label{eq:acr}
 \beta_{IV}(D_1+D_2)
 =&\mathbb{E}[Y(1,1)-Y(1,0)|C_1,C_2]w[C_1,C_2]+\mathbb{E}[Y(1,0)-Y(0,0)|C_1,C_2]w[C_1,C_2]\\ \notag
 +&\mathbb{E}[Y(1,0)-Y(0,0)|C_1,N_2]w[C_1,N_2]+ \mathbb{E}[Y(0,1)-Y(0,0)|N_1,C_2]w[N_1,C_2]\\ \notag
 +&\mathbb{E}[Y(1,1)-Y(0,1)|C_1,A_2]w[C_1,A_2]+\mathbb{E}[Y(1,1)-Y(1,0)|A_1,C_2]w[A_1,C_2],
\end{align}
with $w[G_1,G_2]=\frac{\mathbb{P}[G_1,G_2]}{2\mathbb{P}[C_1,C_2]+\mathbb{P}[C_1,N_2]+\mathbb{P}[N_1,C_2]+\mathbb{P}[C_1,A_2]+\mathbb{P}[A_1,C_2]}$. This results in a causal interpretation of a weighted average of causal effects of receiving one part of the treatment, instead of the causal effect of receiving full treatment. 

The IV estimand $\beta_{IV}(D_1+D_2)$ is related to the average causal response (ACR) as introduced by \cite{angrist1995two}. The ACR is also a weighted average of the effects of unit changes in treatment, but considers a multivalued treatment that does not distinguish between $\{D_1=1,D_2=0\}$ and $\{D_1=0,D_2=1\}$. Without late-adopters who miss the first part of the program, the mover types $\{C_1,A_2\}$ and $\{N_1,C_2\}$ are absent and \eqref{eq:acr} is equal to the ACR. 

Summarising the multivalued treatment by one binary treatment indicator may also lead to violations of the exclusion restriction similar to the ones discussed in Section~\ref{sec:movers}. \citet{andresen2021instrument} show that these violations can also be ruled out by restricting treatment effect heterogeneity or mover types, where movers with a multivalued treatment are defined by the individuals induced by the instrument to change treatment status from and to treatment values that are both below or above the binarisation threshold.

The ACR type estimand in \eqref{eq:acr} considers shifting from no treatment to the first part and from the first part to full treatment, as separate treatment effects for the full compliers. Therefore, it double counts the full compliers in the denominator of the weighting function $w[G_1,G_2]$. An alternative is to combine the two effects for the full compliers and to interpret it as the full treatment effect. \Cref{ass:iv2} allows for the partial identification of this alternative weighted average of LATEs:
\begin{corollary}[Partial identification weighted average of LATEs]\label{corr:partialIV}\ \\
Under \Cref{ass:iv2} a weighted average of LATEs denoted by $\tau$ is bounded by
\begin{align}
\frac{\Delta\mathbb{E}[Y|Z]}{\Delta\mathbb{E}[D_1+D_2|Z]} \leq \tau \leq \frac{\Delta\mathbb{E}[Y|Z]}{\max(\Delta\mathbb{E}[D_1|Z],\Delta\mathbb{E}[D_2|Z],\Delta\mathbb{E}[D_\land|Z],\Delta\mathbb{E}[D_\lor|Z])},
\end{align}
where $\tau$ is defined as
\begin{align}\label{eq:delta}
    \tau = &\mathbb{E}[Y(1,1)-Y(0,0)|C_1,C_2]w[C_1,C_2]+ 
    \mathbb{E}[Y(1,0)-Y(0,0)|C_1,N_2]w[C_1,N_2]+ \\ \notag
    &\,\mathbb{E}[Y(1,1)-Y(0,1)|C_1,A_2]w[C_1,A_2]+ 
    \mathbb{E}[Y(0,1)-Y(0,0)|N_1,C_2]w[N_1,C_2] +\\ \notag
     &\, \mathbb{E}[Y(1,1)-Y(1,0)|A_1,C_2]w[A_1,C_2],
\end{align}
with $w(G_1,G_2)=\frac{\mathbb{P}[G_1,G_2]}{\mathbb{P}[C_1,C_2]+\mathbb{P}[C_1,N_2]+\mathbb{P}[C_1,A_2]+\mathbb{P}[N_1,C_2]+\mathbb{P}[A_1,C_2]}$.
\end{corollary}
The proof is based on \Cref{lemma:fs_rf} and deferred to \Cref{A:proof_corr_partial}. The weights in \eqref{eq:delta} are nonnegative and add up to one. Hence $\tau$ is a convex combination of LATEs and has a causal interpretation. However, this interpretation may not directly be policy relevant as it measures an average across different treatment effects and different groups.  

Intuitively, the treatment indicator with the largest (smallest) first stage estimate may provide a lower (upper) bound on a treatment effect of interest. For instance, \cite{finkelstein2012oregon} apply this intuition to bound the effects of Medicaid insurance. \Cref{corr:partialIV} shows that this intuition does not apply to $\tau$: Instead of bounding $\tau$ from below, the treatment indicator with the largest first stage is an upper bound on $\tau$.

\section{Empirical applications}\label{sec:application}

This section illustrates the empirical relevance of our methods for treatment evaluation. We consider the LAFTE of a health program, a labour program, an educational program, and a development program. Additional details on the empirical specifications are deferred to \Cref{A:app}. This section focuses on the main results following from  \Cref{theorem:late_c} and \Cref{prop:nec_cond_movers} and \ref{prop:nec_cond_double}. \Cref{A:app2} shows additional empirical results, such as the bounds in \Cref{corr:bounded}, which are in general wide and cannot reject that the LAFTEs are equal to zero.

\subsection{The Oregon health insurance experiment}
\label{sec:OHIE}

This application studies the LAFTE of medicaid coverage on health care utilization. Policy makers actively aim to reduce coverage disruptions by continuous enrollment policies. As a recent example, the Families First Coronavirus Recovery Act requires Medicaid programs to keep individuals enrolled for the duration of the public health emergency \citep{sugar2021medicaid}. To better understand the impact of such policies, the effect of continuous Medicaid coverage across the full study period is of particular interest. 

\subsubsection{Context and empirical specification}

In 2008, a group of uninsured low-income adults in Oregon was randomly given the opportunity to apply for Medicaid. The state opened a waiting list for 10,000 Medicaid spots and subsequently drew names by lottery from the 89,924 individuals that placed themselves on this list. The Medicaid program provided comprehensive benefits with no consumer cost sharing. The monthly enrollment premiums ranged from \$0 to \$20 depending on income. 

\cite{finkelstein2012oregon} use the randomized opportunity to apply for Medicaid as an instrument for Medicaid enrollment and analyze the effects on health care utilization, financial strain, and health. The Oregon Health Insurance Experiment (OHIE) was subsequently used in a series of papers to study the effects of Medicaid on other outcomes. For example, \cite{finkelstein2016effect} study the impact on emergency department use over time.

The data in \cite{finkelstein2016effect} contains Medicaid coverage and emergency department use for 24,646 individuals with Portland-area zip codes across four time periods: day 0 to 180, 181 to 360, 361 to 540, and 541 to 720 after lottery notification. Our $D_1$ equals one if an individual was enrolled in Medicaid in the first year after lottery notification, and $D_2$ equals one if an individual was enrolled in the second year after lottery notification. The instrument equals one if an individual was randomly given the opportunity to apply for Medicaid. The binary outcome variable $Y$ equals one if an individual had any emergency department visit during the two years after the lottery notification. 

\subsubsection{Results}

\Cref{tbl:assumptions} reports the estimated necessary conditions for the presence of movers and the validity of \Cref{ass:double_exclusion}. Column (1) and (2) in Panel A show, respectively, the results of a regression from $D_{\lor}-D_2$ and $D_{\land}-D_2$ upon $Z$. First, since both estimates are significantly different from zero at the 1\% level, we reject that movers are absent. Second, since both estimates are positive, we do not reject the necessary conditions of \Cref{ass:double_exclusion}.

\begin{table}[t]
\small
  \begin{center}
    \caption{Necessary conditions for the double exclusion restriction and no movers}
    \label{tbl:assumptions}
\begin{tabular}{l*{5}{c}}
\toprule\toprule
       {\color{white} widercol}             &\multicolumn{1}{c}{(1)}&\multicolumn{1}{c}{(2)}&\multicolumn{1}{c}{(3)}&\multicolumn{1}{c}{(4)}&\multicolumn{1}{c}{(5)}\\
                    &\multicolumn{1}{c}{ $ D_{\lor}-D_{2} $ }&\multicolumn{1}{c}{ $ D_{\land}-D_{2} $ }&\multicolumn{1}{c}{ $ (D_{\lor}-D_{2})Y $ }&\multicolumn{1}{c}{ $ (D_{\land}-D_{2})Y $ }&\multicolumn{1}{c}{ $ D_{2}$ }\\
                    \midrule
                     \multicolumn{6}{c}{Panel A: The Oregon health insurance experiment, $N=24646$} \\
 $ Z $              &       0.078&       0.060&       0.032&       0.029&       0.115\\
                    &     (0.003)&     (0.003)&     (0.002)&     (0.002)&     (0.006)\\
                    \midrule
                     \multicolumn{6}{c}{Panel B: The LinkedIn job opportunities experiment, $N=988$} \\
 $ Z $              &       0.036&       0.042&       0.027&       0.023&       0.315\\
                    &     (0.039)&     (0.011)&     (0.032)&     (0.006)&     (0.077)\\
                    \midrule
                     \multicolumn{6}{c}{Panel C: The elite university education natural experiment, $N=645$} \\
 $ Z $              &       0.047&       0.537&       0.446&       4.772&       0.095\\
                    &     (0.012)&     (0.043)&     (0.116)&     (0.387)&     (0.061)\\
                    \midrule
                    \multicolumn{6}{c}{Panel D: The village-based schools experiment, $N=1181$} \\
 $ Z $              &       0.000&      -0.041&       0.000&       0.016&       0.483\\
                    &         (.)&     (0.008)&         (.)&     (0.006)&     (0.080)\\
       \bottomrule \bottomrule
    \end{tabular}
        \begin{tablenotes}
	\footnotesize
	\item[] Notes: columns (1) to (4) show the estimates for the necessary conditions in \Cref{prop:nec_cond_movers} and \ref{prop:nec_cond_double}, corresponding to \eqref{eq:nC_1}-\eqref{eq:nc4} respectively. Column (5) shows the estimate of a regression from $D_{2}$ upon $Z$. The standard errors are in parentheses. The panels correspond to the empirical applications in \Cref{sec:application}, where $N$ indicates the number of observations.
\end{tablenotes}
  \end{center}
\end{table}

Under the double exclusion restriction, column (1) implies that $\widehat{\mathbb{P}[C_1,N_2]}=0.078$. This group includes movers who drop out of treatment. Since individuals had to recertify eligibility every six months, they could have lost coverage due to, for instance, income fluctuations. Column (2) implies that $\widehat{\mathbb{P}[C_1,A_2]}=0.060$. These movers likely used the opportunity to gain eligibility in the second time period: 14 months after the lottery, Oregon took lottery draws from a new Medicaid waiting list. Column (5) reports the results of a regression from $D_2$ upon $Z$. Under the double exclusion restriction this estimate equals $\widehat{\mathbb{P}[C_1,C_2]}=0.115$, and hence the movers compose a larger proportion of the data than the full compliers.

The double exclusion restriction imposes that $\mathbb{P}[N_1,C_2]=\mathbb{P}[A_1,C_2]=0$. The first mover type is likely absent since individuals with $Z=1$ had to apply for Medicaid within 45 days after the state made them aware of the lottery draw. Individuals that did not apply during this window could not apply after. In general, \cite{finkelstein2012oregon} describe that the mechanisms to obtain Medicaid coverage are limited to the initial lottery and the lottery 14 months later. They also show that only 2\% of the individuals with $Z=0$ obtained Medicaid coverage up unto one year after the lottery. This may explain the absence of the second mover type.  

\begin{figure}[t]
\centering
\caption{Estimated bounds on the LAFTE}
\label{fig:bounds} 
\includegraphics[width=1\textwidth]{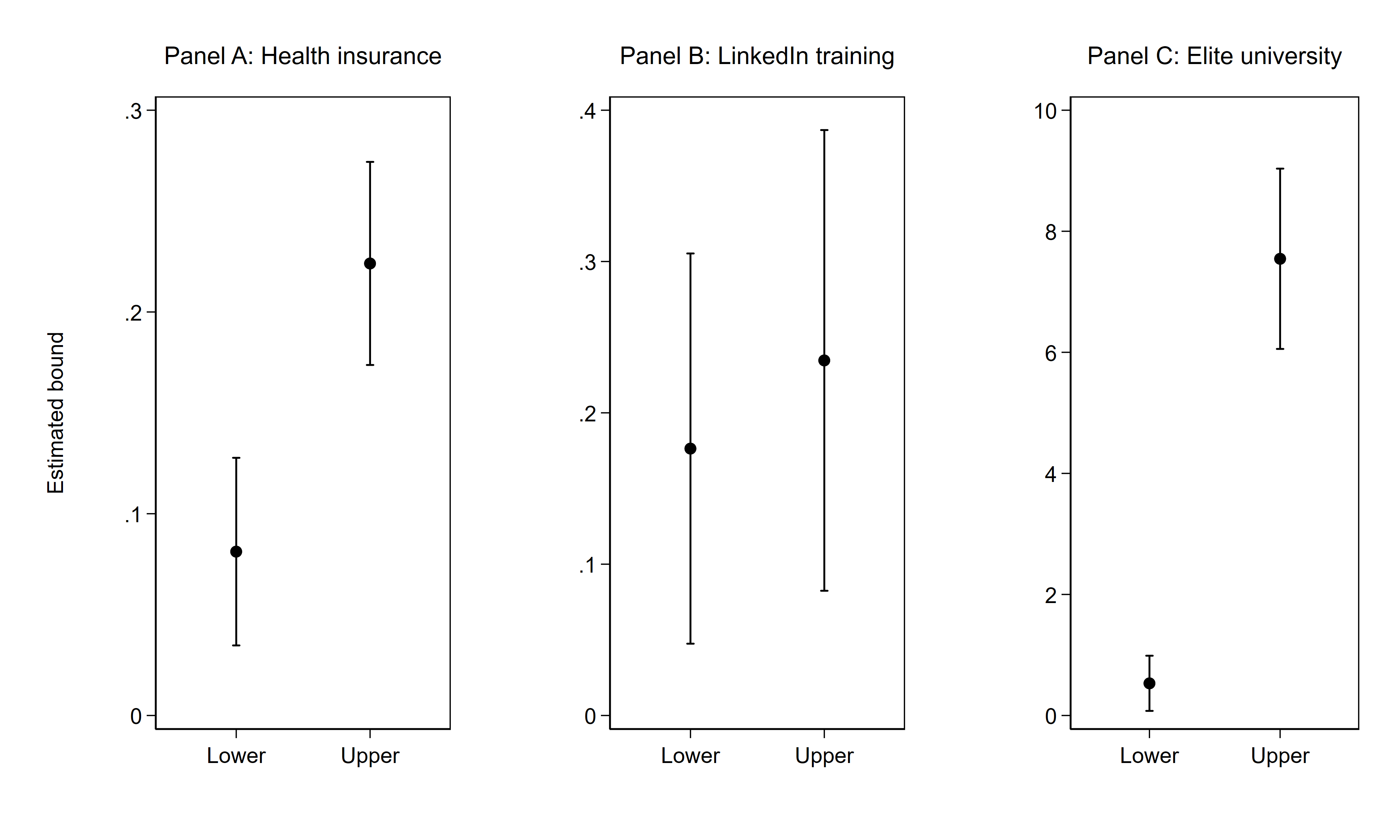} 
\caption*{ \footnotesize
Notes: this figure shows the lower and upper bound from \Cref{theorem:late_c} for the LAFTE. The lines are 95\% confidence intervals. The precise estimates and standard errors are reported in \Cref{tbl:bounds} of \Cref{A:app2}.
}
\end{figure}

\Cref{fig:bounds} shows the estimated bounds for the LAFTE from \Cref{theorem:late_c}. Panel A identifies a statistically significant LAFTE: The 95\% confidence interval of the lower bound, $[0.035,0.128]$, does not include zero. The bounds rely on the assumption that the effect of Medicaid in the second year on emergency department use is non-negative (MTR), and that always takers and compliers of Medicaid in the second year are not less likely to use the emergency department than never takers (MTS). Based on these bounds, we conclude that Medicaid enrollment during both years increases the probability of an emergency department visit between 0.081 and 0.224 percentage points for the individuals induced to enroll by the lottery. 

The estimated bounds imply that the LAFTE may be up to three times as large as the LATE of Medicaid enrollment in the first period. Hence, policy makers may expect larger treatment effects in combination with continuous enrollment requirements, which may inform the design of future Medicaid enrollment policies.

\subsection{The LinkedIn job opportunities experiment}

This application studies the LAFTE of a LinkedIn training on employment. Job market training programs, such as a training that stimulates LinkedIn usage, are often characterized by high incidence of control group substitution and treatment group dropout \citep{heckman2000substitution}. Since the programs are carefully designed to receive and follow in full, the treatment effect of LinkedIn usage for a longer period should be of particular interest to policy makers. 

\subsubsection{Context and empirical specification}

\cite{wheeler2022linkedin} run and evaluate a randomly assigned program that trains work seekers to join and use LinkedIn. Their study sample includes 30 cohorts from existing job readiness training programs in four large South African cities. They randomly assign 15 cohorts to four hours of LinkedIn training during their job readiness training program. The intervention trains participants, among others, to open accounts, build their profiles, and search and apply for jobs. The study examines the effect of the LinkedIn training on a range of outcomes. Although the experiment is not specifically designed to identify the causal effect of LinkedIn usage, one of the analysis uses the random assignment to LinkedIn training as an instrument for LinkedIn usage to study its effect on employment.

The data includes LinkedIn usage and employment outcomes across three time periods: directly, six months, and twelve months after the end of the job readiness training program. Our $D_1$ equals one if a participant has a Linkedin account at the end of the job training and $D_2$ equals one if a participant has a LinkedIn account six months later. The instrument $Z$ equals one if the participant is randomly assigned to the LinkedIn training. The binary outcome variable $Y$ equals one if the participant is employed twelve months after the program ended. The estimation sample includes 988 of the 1,638 participants across the 30 cohorts. 
 
\subsubsection{Results}

The estimated necessary conditions in column (1) and (2) of Panel B in \Cref{tbl:assumptions} are positive, and the latter estimate is also significantly different from zero at the 1\% level. Hence, we reject that movers are absent and do not reject the necessary conditions of \Cref{ass:double_exclusion}.

Under the double exclusion restriction, column (1) implies that $\widehat{\mathbb{P}[C_1,N_2]}=0.036$, which is not significantly different from zero. This estimate refers to individuals that were induced to open a LinkedIn account by the LinkedIn training but deleted it after the job training program. Column (2) implies that $\widehat{\mathbb{P}[C_1,A_2]}=0.042$. This suggests that several treated individuals would have also opened a LinkedIn account after the job training program if assigned to the control. Hence, the LinkedIn training only accelerated them to create an account. Column (5) implies that the two groups of movers are small compared to the full compliers.

The double exclusion restriction imposes that $\mathbb{P}[N_1,C_2]=\mathbb{P}[A_1,C_2]=0$. Since the treatment group was incentivized to immediately open a LinkedIn account in the first week of the job training program, the first mover type can be absent. However, if the LinkedIn training succeeds in explaining the benefits of LinkedIn in the long term, both mover types could be present. Hence, even though they are not detected by the necessary conditions, this empirical setting may include movers that violate the double exclusion restriction.

Panel B of \Cref{fig:bounds} shows a statistically significant LAFTE: The 95\% confidence interval of the lower bound, $[0.047,0.305]$, does not include zero. The bounds rely on the assumption that the effect of a LinkedIn account six months after the job training program on employment is non-negative (MTR), and that always takers and compliers of a LinkedIn account six monthts after the job training are not less likely to find a job than never takers (MTS). Based on these bounds, we conclude that having a Linkedin account in both time periods increases the probability to find a job between 0.176 and 0.235 percentage points for individuals induced to open an account due to the LinkedIn training. The estimated bounds imply that the LAFTE is similar to the LATE of the first part of the LinkedIn training, which suggests that having a Linkedin account is most effective shortly after the job training program.

\subsection{The elite university education natural experiment}

This application studies the LAFTE of a university education on income. School curricula, including the design and sequencing of learning content, are developed to receive in full. Moreover, students enroll in a program with the aim to graduate, which makes the LAFTE a valuable piece of information for their study choice. Hence, for both university educators and students the effect of following a full university program is of interest. 

\subsubsection{Context and empirical specification}

\cite{anelli2020returns} studies the returns to a selective, expensive, and private university offering business, economics, and law degrees in a large Italian city. Admission to the elite university is based on a uni-dimensional application score. Every year, the number of admitted students is fixed, and the university strictly offers admission to the students with the highest application score. This procedure generates a cutoff in the application score, where students above the cutoff are offered admission. A score above the cutoff is used as an instrument for ever enrolled at the elite university to study the effect of ever enrolled on income.

The main sample is restricted to the elite university applicants between 1995 and 2000. For 645 of these applicants the dataset contains information on the application score, ever being enrolled at the elite university between 1995 and 2005, graduation from any university up unto the year 2005, and yearly income in the year 2005. If an applicant retook the admission test, the score refers to the first observed application. 

Our $D_1$ equals one if an applicant was ever enrolled at the elite university between 1995 and 2005. This is the treatment variable in \cite{anelli2020returns}. Our $D_2$ equals one if an applicant graduated from any university up unto the year 2005. The dataset does not contain information on whether the applicant graduated from the elite university. The instrument $Z$ equals one if the applicant scores above to application score cutoff. The outcome variable $Y$ is the logarithm of taxable income in 2005, measured before taxes and after deductions. 

\subsubsection{Results}

The estimated necessary conditions in column (1) and (2) of Panel C in \Cref{tbl:assumptions} are positive and significantly different from zero at the 1\% level. Hence, we can reject that movers are absent and do not reject the necessary conditions of \Cref{ass:double_exclusion}.

Under the double exclusion restriction, column (1) implies that $\widehat{\mathbb{P}[C_1,N_2]}=0.047$. This mover type may refer to students who were induced to enroll at the elite university but never graduate. Column (2) implies that $\widehat{\mathbb{P}[C_1,A_2]}=0.537$. Recall that our $D_2$ equals one if a student graduated from any university, not just the elite institution, and so this group is relatively large because students with $Z=0$ may graduate from other universities. 

The double exclusion restriction imposes that $\mathbb{P}[N_1,C_2]=\mathbb{P}[A_1,C_2]=0$. The first mover type is absent if university graduation for the never takers of elite university enrollment is not affected by scoring above or below the cutoff. This group of movers could be present if scoring above the cutoff has some positive psychological effect that persists even if a student does not enroll at the elite institution. The large estimate in column (2) of \Cref{tbl:assumptions} suggests that such psychological effects are close to zero. The second mover type is likely to be absent since \cite{anelli2020returns} reports that across all application rounds only 23 students retook the admission test after  scoring below the cutoff. Hence the possible number of always takers with enrollment in the elite university is small.

Panel C of \Cref{fig:bounds} shows a statistically significant LAFTE: the estimated bounds equal 0.532 and 7.545, and the 95\% confidence interval of the lower bound, $[0.076, 0.988]$, does not include zero. The bounds rely on the assumption that the effect of university graduation on income is non-negative, and that always takers and compliers of university graduation have higher wages than never takers. The upper bound is large due to the large group of movers $\{C_1,A_2\}$. Based on the bounds, we conclude that the full treatment of enrolling at the elite and graduating at any university increases annual income between 53 and 750 log points for the individuals induced to enroll by scoring above the cutoff. The estimated bounds imply that the LAFTE may be much larger than the LATE of university enrollment. This suggests that consecutive parts of a university degree are complements and students may particularly benefit from fully completing a degree.

\subsection{The village-based schools experiment}

This application aims to study the LAFTE of a development program. The implementation of development programs often requires careful design, planning, and coordination with local partners, and can be expensive \citep{duflo2007using}. Hence, the full treatment effect is of interest to the stakeholders. 

\subsubsection{Context and empirical specification}

\cite{burde2013bringing} conduct and evaluate the randomized opening of village-based schools in Afghanistan, where primary-school participation rates are low. Their study sample includes 31 villages in rural Afghanistan. In the summer of 2007, they randomly opened schools in 13 of these villages. Village-based schools are public schools that are designed to deliver the official national curriculum to children living in close proximity to the school. They use the random assignment of village-based schools as an instrument for school enrollment to estimate the effect of enrollment on academic performance.  

Their household survey data contains information on school enrollment and math and language test scores in two time periods: four months after and eight months after opening the village-based schools. Our $D_1$ equals one if a child was enrolled in school four months after the opening and $D_2$ equals one if a child was enrolled in school eight months after the opening. The binary instrument $Z$ equals one if a child lives in a village that randomly received a school. The outcome variable $Y$ is the standardized test score eight months after the opening of the schools. The final estimation sample includes 1,181 children, of in total 1,490 school-age children across the 31 villages.

\subsubsection{Results}

Panel D in \Cref{tbl:assumptions} shows the estimated necessary conditions. Since the observed
$D_2=1$ if $D_1=1$, the variable $D_{\lor}-D_2$ is zero for each individual, and hence the estimates in column (1) and (3) are zero. This indicates the absence of the mover types $\{C_1,N_2\}$ and $\{A_1,C_2\}$. The estimate in column (2) is negative and statistically significant at the 1\% level. This estimate implies that $\mathbb{P}[N_1,C_2]>0$, which violates the double exclusion restriction. This mover type may arise if households need time to change their beliefs towards allowing their children to enroll in school. \cite{burde2013bringing} discuss that the conservative beliefs in Afghanistan may be an impediment for school enrollment, in particular for girls.

\Cref{A:app2} shows the estimates for the parameters that can be identified when the double exclusion restriction is violated: the ACR type estimand in \eqref{eq:acr} and the weighted average of LATEs in \Cref{corr:partialIV}. To conclude, this application shows that it depends on the empirical context whether the double exclusion restriction holds, and that the testable necessary conditions are able to help researchers detect violations of this assumption.

\section{Conclusion}\label{sec:conclusion}

Policy evaluation has to deal with individuals that only take a subset of a treatment program. This poses the question: Which treatment effect is of interest to the policymaker? For instance, are the outcomes of the individuals receiving the full treatment or any individual that came in contact with the treatment of interest? Does the control group consists of individuals who obtained zero treatment or did not complete the full treatment? This paper defines the effect of full treatment versus no treatment at all as the parameter of interest.  

We develop an IV framework in the presence of individuals that are induced by the instrument to obtain only a subset of the treatment, and refer to these individuals as movers. We show that these movers violate the exclusion restriction in the standard IV framework, that necessary conditions on the presence of movers are testable, and that under a double exclusion restriction the average treatment effect of receiving full treatment for the individuals induced into treatment by the instrument is partially identifiable. We refer to this treatment effect as the local average full treatment effect (LAFTE).

We study the LAFTE in empirical applications from four different fields of applied economics. Our methods find  evidence for the presence of movers in all four studies. These types of movers align with the intuition provided by the setting at hand. In three of these applications, the necessary conditions for the double exclusion restriction cannot be rejected, and the bounds identify a statistically significant LAFTE. We discuss the potential policy implications within the empirical context of the applications.

We allow the instrument to induce individuals into dropping out or late-adoption of treatment by taking into account two treatment parts. In settings in which the instrument additionally affects dropping out or late-adoption across more treatment parts, the exclusion restriction in our IV framework may also be violated. In case these settings include the observed treatment status across these treatment parts, our framework can be extended to more specific mover types that also take these violations into account.

\linespread{1.00}
\normalsize
\addcontentsline{toc}{section}{References}
\bibliographystyle{chicagoa}
\bibliography{Library}
\normalsize

\newpage

\appendix

\section{\texorpdfstring{Proof \Cref{lemma:fs_rf}}{proof}}\label{A:proof_lem1}

The first stage with $D=D_1$ equals
\begin{align}\label{eq:fspo_d1}
    \Delta\mathbb{E}[D_1|Z] =& \mathbb{E}[D_1(1)|Z=1]-\mathbb{E}[D_1(0)|Z=0] = 
    \mathbb{E}[D_1(1)-D_1(0)]=\mathbb{P}[D_1(1)-D_1(0)=1]\\ \notag
    =&\mathbb{P}[C_1]= \mathbb{P}[C_1,D_2(1,1)-D_2(0,0)=1] +\mathbb{P}[C_1,D_2(1,1)=D_2(0,0)=0]+  \\ \notag 
    &\mathbb{P}[C_1,D_2(1,1)=D_2(0,0)=1] \\ \notag
    =&\mathbb{P}[C_1,C_2]+\mathbb{P}[C_1,N_2]+\mathbb{P}[C_1,A_2],
\end{align}
using \eqref{eq:po_d1}, \Cref{ass:iv2}.1 and \ref{ass:iv2}.3, respectively.

The first stage with $D=D_2$ equals
\begin{align}\label{eq:fspo_d2}
    \Delta\mathbb{E}[D_2|Z]  =& \mathbb{E}[D_2(1,1)D_1+D_2(1,0)(1-D_1)|Z=1]-\\ \notag
    &\mathbb{E}[D_2(0,1)D_1+D_2(0,0)(1-D_1)|Z=0] \\ \notag
    =& \mathbb{E}[D_2(1,1)D_1(1)+D_2(1,0)(1-D_1(1))|Z=1]-\\ \notag
    &\mathbb{E}[D_2(0,1)D_1(0)+D_2(0,0)(1-D_1(0))|Z=0] \\ \notag
    =& \mathbb{E}[D_2(1,1)D_1(1)+D_2(1,0)(1-D_1(1))-D_2(0,1)D_1(0)-D_2(0,0)(1-D_1(0))]\\ \notag
    =& \mathbb{E}[D_2(1,1)-D_2(0,0)|D_1(1)-D_1(0)=1]\mathbb{P}[D_1(1)-D_1(0)=1]+\\ \notag
    &\mathbb{E}[D_2(1,0)-D_2(0,0)|D_1(1)=D_1(0)=0]\mathbb{P}[D_1(1)=D_1(0)=0]+\\ \notag
    &\mathbb{E}[D_2(1,1)-D_2(0,1)|D_1(1)=D_1(0)=1]\mathbb{P}[D_1(1)=D_1(0)=1]\\ \notag
    =& \mathbb{P}[D_2(1,1)-D_2(0,0)=1|D_1(1)-D_1(0)=1]\mathbb{P}[D_1(1)-D_1(0)=1]+\\ \notag
    &\mathbb{P}[D_2(1,0)-D_2(0,0)=1|D_1(1)=D_1(0)=0]\mathbb{P}[D_1(1)=D_1(0)=0]+\\ \notag
    &\mathbb{P}[D_2(1,1)-D_2(0,1)=1|D_1(1)=D_1(0)=1]\mathbb{P}[D_1(1)=D_1(0)=1]\\ \notag
=& \mathbb{P}[D_2(1,1)-D_2(0,0)=1,D_1(1)-D_1(0)=1]+\\ \notag
    &\mathbb{P}[D_2(1,0)-D_2(0,0)=1,D_1(1)=D_1(0)=0]+\\ \notag
    &\mathbb{P}[D_2(1,1)-D_2(0,1)=1,D_1(1)=D_1(0)=1]\\ \notag
    =& \mathbb{P}[C_1,C_2]+\mathbb{P}[N_1,C_2]+\mathbb{P}[A_1,C_2],
\end{align}
using \eqref{eq:po_d2}, \eqref{eq:po_d1}, \Cref{ass:iv2}.1 and \ref{ass:iv2}.3, respectively.

The first stage with $D=D_\land=D_1D_2$ can be rewritten using \eqref{eq:po_d1} and \eqref{eq:po_d2}:
\begin{align}\label{eq:po_dand}
    D_\land
    =& [D_1(1)Z+D_1(0)(1-Z)]\times [D_2(1,1)(D_1(1)Z+D_1(0)(1-Z))Z+ \\ \notag
    &D_2(1,0)(1-D_1(1)Z-D_1(0)(1-Z))Z+ \\ \notag
    &D_2(0,1)(D_1(1)Z+D_1(0)(1-Z))(1-Z)+ \\ \notag
    &D_2(0,0)(1-D_1(1)Z-D_1(0)(1-Z))(1-Z)] \\ \notag
    =& [D_1(1)Z+D_1(0)(1-Z)]\times [D_2(1,1)D_1(1)Z+D_2(1,0)(1-D_1(1))Z+ \\ \notag
    &D_2(0,1)D_1(0)(1-Z)+D_2(0,0)(1-D_1(0))(1-Z)] \\ \notag
    =& D_1(1)D_2(1,1)Z+D_1(0)D_2(0,1)(1-Z),
\end{align}
so that
\begin{align}\label{eq:fspo_dand}
    \Delta\mathbb{E}[D_\land|Z] =& \mathbb{E}[D_1(1)D_2(1,1)|Z=1]-\mathbb{E}[D_1(0)D_2(0,1)|Z=0] \\ \notag
    =& \mathbb{E}[D_1(1)D_2(1,1)-D_1(0)D_2(0,1)]\\ \notag
    =& \mathbb{P}[D_1(1)D_2(1,1)-D_1(0)D_2(0,1)=1]\\ \notag
    =    &\mathbb{P}[D_1(1)-D_1(0)=1,D_2(1,1)-D_2(0,1)=1]+\\ \notag
    & \mathbb{P}[D_1(1)-D_1(0)=1,D_2(1,1)=D_2(0,1)=1]+\\ \notag
    &\mathbb{P}[D_1(1)=D_1(0)=1,D_2(1,1)-D_2(0,1)=1]\\ \notag
    =& \mathbb{P}[C_1,C_2]+\mathbb{P}[C_1,A_2]+\mathbb{P}[A_1,C_2],
\end{align}
using \Cref{ass:iv2}.1 and \ref{ass:iv2}.3, respectively.

The first stage with $D=D_\lor=D_1+D_2-D_1D_2$ equals
\begin{align}\label{eq:fspo_dor}
    \Delta\mathbb{E}[D_\lor|Z]=&\Delta\mathbb{E}[D_1|Z]+\Delta\mathbb{E}[D_2|Z]-\Delta\mathbb{E}[D_{\land}|Z] \\ \notag
    =& \mathbb{P}[C_1,C_2]+\mathbb{P}[C_1,N_2]+\mathbb{P}[N_1,C_2].
\end{align}

For the reduced form, we use \Cref{ass:iv2}.4 and rewrite \eqref{eq:Yd} to
\begin{align}
    Y=&[Y(1,1)-Y(0,1)-Y(1,0)+Y(0,0)]D_1D_2+\\ \notag
    &[Y(1,0)-Y(0,0)]D_1+[Y(0,1)-Y(0,0)]D_2+Y(0,0), 
\end{align}
to write
\begin{align}
    \Delta\mathbb{E}[Y|Z]=&\Delta\mathbb{E}[[Y(1,1)-Y(0,1)-Y(1,0)+Y(0,0)]D_1D_2|Z]+\\ \notag
    &\Delta\mathbb{E}[[Y(1,0)-Y(0,0)]D_1|Z]+\Delta\mathbb{E}[[Y(0,1)-Y(0,0)]D_2|Z]\\ \notag
    =&\mathbb{E}[(Y(1,1)-Y(0,1)-Y(1,0)+Y(0,0))D_1D_2|Z=1]-  \\ \notag 
    &\mathbb{E}[(Y(1,1)-Y(0,1)-Y(1,0)+Y(0,0))D_1D_2|Z=0]+\\ \notag
    &\mathbb{E}[(Y(1,0)-Y(0,0))D_1|Z=1]-\mathbb{E}[(Y(1,0)-Y(0,0))D_1|Z=0]+ \\ \notag
    &\mathbb{E}[(Y(0,1)-Y(0,0))D_2|Z=1]-\mathbb{E}[(Y(0,1)-Y(0,0))D_2|Z=0].
\end{align}
It follows from the results in \eqref{eq:fspo_d1}, \eqref{eq:fspo_d2}, and \eqref{eq:fspo_dand} that 
\begin{align}
    \Delta\mathbb{E}[Y|Z]=& \mathbb{E}[Y(1,1)-Y(0,1)-Y(1,0)+Y(0,0)|\{C_1,C_2\},\{C_1,A_2\},\{A_1,C_2\}]\times\\ \notag 
    &\mathbb{P}[\{C_1,C_2\},\{C_1,A_2\},\{A_1,C_2\}]+\\ \notag
    &\mathbb{E}[Y(1,0)-Y(0,0)|\{C_1,C_2\},\{C_1,N_2\},\{C_1,A_2\}]\times\\ \notag 
    &\mathbb{P}[\{C_1,C_2\},\{C_1,N_2\},\{C_1,A_2\}]+\\ \notag
    &\mathbb{E}[Y(0,1)-Y(0,0)|\{C_1,C_2\},\{N_1,C_2\},\{A_1,C_2\}] \times \\ \notag 
    &\mathbb{P}[\{C_1,C_2\},\{N_1,C_2\},\{A_1,C_2\}]\\ \notag
        =&\mathbb{E}[Y(1,1)-Y(0,1)-Y(1,0)+Y(0,0)|C_1,C_2]\mathbb{P}[C_1,C_2]+\\ \notag
    &\mathbb{E}[Y(1,1)-Y(0,1)-Y(1,0)+Y(0,0)|C_1,A_2]\mathbb{P}[C_1,A_2]+\\ \notag
    &\mathbb{E}[Y(1,1)-Y(0,1)-Y(1,0)+Y(0,0)|A_1,C_2]\mathbb{P}[A_1,C_2]+\\ \notag
    &\mathbb{E}[Y(1,0)-Y(0,0)|C_1,C_2]\mathbb{P}[C_1,C_2]+
    \mathbb{E}[Y(1,0)-Y(0,0)|C_1,N_2]\mathbb{P}[C_1,N_2]+\\ \notag
    &\mathbb{E}[Y(1,0)-Y(0,0)|C_1,A_2]\mathbb{P}[C_1,A_2]+
    \mathbb{E}[Y(0,1)-Y(0,0)|C_1,C_2]\mathbb{P}[C_1,C_2]+\\ \notag
    &\mathbb{E}[Y(0,1)-Y(0,0)|N_1,C_2]\mathbb{P}[N_1,C_2]+
    \mathbb{E}[Y(0,1)-Y(0,0)|A_1,C_2]\mathbb{P}[A_1,C_2]\\ \notag
      = &\mathbb{E}[Y(1,1)-Y(0,0)|C_1,C_2]\mathbb{P}[C_1,C_2]+
    \mathbb{E}[Y(1,0)-Y(0,0)|C_1,N_2]\mathbb{P}[C_1,N_2]+\\ \notag
    &\mathbb{E}[Y(1,1)-Y(0,1)|C_1,A_2]\mathbb{P}[C_1,A_2]+
    \mathbb{E}[Y(0,1)-Y(0,0)|N_1,C_2]\mathbb{P}[N_1,C_2]+\\ \notag
    &\mathbb{E}[Y(1,1)-Y(1,0)|A_1,C_2]\mathbb{P}[A_1,C_2]. 
\end{align}

\section{\texorpdfstring{Proof \Cref{cor:homo_movers}}{proof}}\label{A:proof_cor2}

For an instrumental variable analysis with treatment indicator $D=D_1$, it follows from \Cref{lemma:fs_rf} that the first stage and reduced form is equal to 
\begin{align}
    \Delta\mathbb{E}[D_1|Z] =&\mathbb{P}[C_1,C_2]+\mathbb{P}[C_1,N_2]+\mathbb{P}[C_1,A_2]\\ \notag
    \Delta\mathbb{E}[Y|Z] = &\mathbb{E}[Y(1,1)-Y(0,0)|C_1,C_2]\mathbb{P}[C_1,C_2]+
    \mathbb{E}[Y(1,0)-Y(0,0)|C_1,N_2]\mathbb{P}[C_1,N_2]+\\ \notag
    &\mathbb{E}[Y(1,1)-Y(0,1)|C_1,A_2]\mathbb{P}[C_1,A_2]+
    \mathbb{E}[Y(0,1)-Y(0,0)|N_1,C_2]\mathbb{P}[N_1,C_2]+\\ \notag
    &\mathbb{E}[Y(1,1)-Y(1,0)|A_1,C_2]\mathbb{P}[A_1,C_2].
\end{align}
We write the reduced form as,
\begin{align}
     \Delta\mathbb{E}[Y|Z] = &\mathbb{E}[Y(1,1)-Y(0,0)|C_1,C_2]\mathbb{P}[C_1,C_2]+\mathbb{E}[Y(1,0)-Y(0,0)|C_1,N_2]\mathbb{P}[C_1,N_2]+\\ \notag
    &\mathbb{E}[Y(1,1)-Y(0,1)|C_1,A_2]\mathbb{P}[C_1,A_2]+\\ \notag
    &\mathbb{E}[Y(1,1)-Y(0,0)|N_1,C_2]\mathbb{P}[N_1,C_2]-\mathbb{E}[Y(1,1)-Y(0,1)|N_1,C_2]\mathbb{P}[N_1,C_2]+\\ \notag
    &\mathbb{E}[Y(1,1)-Y(0,0)|A_1,C_2]\mathbb{P}[A_1,C_2]-\mathbb{E}[Y(1,0)-Y(0,0)|A_1,C_2]\mathbb{P}[A_1,C_2],
\end{align}
and substitute in the homogeneity assumptions for $D=D_1$ in \Cref{tab:homo_movers}:
\begin{align}
    \Delta\mathbb{E}[Y|Z] = &\mathbb{E}[Y(1,1)-Y(0,0)|C_1,C_2]\mathbb{P}[C_1,C_2]+\mathbb{E}[Y(1,1)-Y(0,0)|C_1,N_2]\mathbb{P}[C_1,N_2]+\\ \notag
    &\mathbb{E}[Y(1,1)-Y(0,0)|C_1,A_2]\mathbb{P}[C_1,A_2]+\\ \notag
    &\mathbb{E}[Y(1,1)-Y(0,0)|N_1,C_2]\mathbb{P}[N_1,C_2]-\mathbb{E}[Y(1,1)-Y(0,0)|N_1,C_2]\mathbb{P}[N_1,C_2]+\\ \notag
    &\mathbb{E}[Y(1,1)-Y(0,0)|A_1,C_2]\mathbb{P}[A_1,C_2]-\mathbb{E}[Y(1,1)-Y(0,0)|A_1,C_2]\mathbb{P}[A_1,C_2]\\ \notag
    =& \mathbb{E}[Y(1,1)-Y(0,0)|C_1,C_2]\mathbb{P}[C_1,C_2]+
    \mathbb{E}[Y(1,1)-Y(0,0)|C_1,N_2]\mathbb{P}[C_1,N_2]+\\ \notag
    &\mathbb{E}[Y(1,1)-Y(0,0)|C_1,A_2]\mathbb{P}[C_1,A_2].
\end{align}
Then we can write
\begin{align}
\beta_{IV}(D_1)=\frac{\Delta\mathbb{E}[Y|Z]}{\Delta\mathbb{E}[D_1|Z]}=&\mathbb{E}[Y(1,1)-Y(0,0)|\{C_1,C_2\},\{C_1,N_2\},\{C_1,A_2\}].
\end{align}
The proofs for $D=D_2,D_\land,D_\lor$ use the same strategy.

\begin{table}[t]
\footnotesize
    \caption{Treatment effect homogeneity assumptions for movers}
    \label{tab:homo_movers}
\begin{tabular}{ l l l} 
 \toprule\toprule
 $D$ & Homogeneity assumption & IV estimand  \\
 \midrule
 $D_1$ &  $\mathbb{E}[Y(1,1)-Y(0,0)|C_1,N_2]=\mathbb{E}[Y(1,0)-Y(0,0)|C_1,N_2]$ & $\mathbb{E}[Y(1,1)-Y(0,0)|\{C_1,C_2\},\{C_1,N_2\},\{C_1,A_2\}]$\\
  & $\mathbb{E}[Y(1,1)-Y(0,0)|C_1,A_2]=\mathbb{E}[Y(1,1)-Y(0,1)|C_1,A_2]$  & \\
   & $\mathbb{E}[Y(1,1)-Y(0,0)|N_1,C_2]=\mathbb{E}[Y(1,1)-Y(0,1)|N_1,C_2]$  & \\
    & $\mathbb{E}[Y(1,1)-Y(0,0)|A_1,C_2]=\mathbb{E}[Y(1,0)-Y(0,0)|A_1,C_2]$ & \\
    & &\\
 $D_2$ & $\mathbb{E}[Y(1,1)-Y(0,0)|C_1,N_2]=\mathbb{E}[Y(1,1)-Y(1,0)|C_1,N_2]$  & $\mathbb{E}[Y(1,1)-Y(0,0)|\{C_1,C_2\},\{N_1,C_2\},\{A_1,C_2\}]$\\
    & $\mathbb{E}[Y(1,1)-Y(0,0)|C_1,A_2]=\mathbb{E}[Y(0,1)-Y(0,0)|C_1,A_2]$  & \\
   & $\mathbb{E}[Y(1,1)-Y(0,0)|N_1,C_2]=\mathbb{E}[Y(0,1)-Y(0,0)|N_1,C_2]$  & \\
    & $\mathbb{E}[Y(1,1)-Y(0,0)|A_1,C_2]=\mathbb{E}[Y(1,1)-Y(1,0)|A_1,C_2]$ & \\
    & &\\
 $D_\land$ & $\mathbb{E}[Y(1,1)-Y(0,0)|C_1,N_2]=\mathbb{E}[Y(1,1)-Y(1,0)|C_1,N_2]$  & $\mathbb{E}[Y(1,1)-Y(0,0)|\{C_1,C_2\},\{C_1,A_2\},\{A_1,C_2\}]$\\
    & $\mathbb{E}[Y(1,1)-Y(0,0)|C_1,A_2]=\mathbb{E}[Y(1,1)-Y(0,1)|C_1,A_2]$  & \\
   & $\mathbb{E}[Y(1,1)-Y(0,0)|N_1,C_2]=\mathbb{E}[Y(1,1)-Y(0,1)|N_1,C_2]$ & \\
    & $\mathbb{E}[Y(1,1)-Y(0,0)|A_1,C_2]=\mathbb{E}[Y(1,1)-Y(1,0)|A_1,C_2]$ & \\
    & &\\
 $D_\lor$ & $\mathbb{E}[Y(1,1)-Y(0,0)|C_1,N_2]=\mathbb{E}[Y(1,0)-Y(0,0)|C_1,N_2]$  & $\mathbb{E}[Y(1,1)-Y(0,0)|\{C_1,C_2\},\{C_1,N_2\},\{N_1,C_2\}]$\\
    & $\mathbb{E}[Y(1,1)-Y(0,0)|C_1,A_2]=\mathbb{E}[Y(0,1)-Y(0,0)|C_1,A_2]$  & \\
   & $\mathbb{E}[Y(1,1)-Y(0,0)|N_1,C_2]=\mathbb{E}[Y(0,1)-Y(0,0)|N_1,C_2]$  & \\
    & $\mathbb{E}[Y(1,1)-Y(0,0)|A_1,C_2]=\mathbb{E}[Y(1,0)-Y(0,0)|A_1,C_2]$ & \\
 \bottomrule\bottomrule
\end{tabular}   
        \begin{tablenotes}
	\footnotesize
	\item[] Notes: this table provides an overview of the homogeneity assumptions in column (2) that are required for the IV estimand in Equation~\eqref{eq:IV} to identify the causal parameter in column (3) when the treatment indicator in column (1) is used as endogenous treatment variable. 
        \end{tablenotes}
\end{table}

\section{Proof \texorpdfstring{\Cref{lemma:fs_rf_double}}{proof}}\label{A:proof_lem3}

Since \Cref{ass:double_exclusion} only applies to $D_2$, $\Delta\mathbb{E}[D_1|Z]$ is equal to \eqref{eq:fspo_d1}:
\begin{align}
    \Delta\mathbb{E}[D_1|Z] =& \mathbb{P}[C_1]= \mathbb{P}[C_1,D_2(1)-D_2(0)=1] +\mathbb{P}[C_1,D_2(1)=D_2(0)=0]+  \\ \notag 
    &\mathbb{P}[C_1,D_2(1)=D_2(0)=1] \\ \notag
    =&\mathbb{P}[C_1,C_2]+\mathbb{P}[C_1,N_2]+\mathbb{P}[C_1,A_2],
\end{align}
using \Cref{ass:iv2}.1 and \ref{ass:iv2}.3.

\Cref{ass:double_exclusion} allows us to link observed to potential outcomes for $D_2$ as follows,
\begin{align}
    D_2=&D_2(1)D_1+D_2(0)(1-D_1)\\ \notag
       =&D_2(1)[D_1(1)Z+D_1(0)(1-Z)]+D_2(0)[1-D_1(1)Z-D_1(0)(1-Z)]\\ \notag
       =&D_2(0)+(D_2(1)-D_2(0))(D_1(1)Z+D_1(0)(1-Z)), 
\end{align}
so that the first stage with $D=D_2$ equals 
\begin{align}\label{eq:fspo_d2_double}
\Delta\mathbb{E}[D_2|Z]  =&\mathbb{E}[D_2(0)+(D_2(1)-D_2(0))(D_1(1)Z+D_1(0)(1-Z))|Z=1]-\\ \notag
    &\mathbb{E}[D_2(0)+(D_2(1)-D_2(0))(D_1(1)Z+D_1(0)(1-Z))|Z=0] \\ \notag
   = &\mathbb{E}[(D_1(1)-D_1(0))(D_2(1)-D_2(0))]
     = \mathbb{P}[(D_1(1)-D_1(0))(D_2(1)-D_2(0))=1]\\ \notag
    =& \mathbb{P}[D_1(1)-D_1(0)=1,D_2(1)-D_2(0)=1] \\ \notag
     =& \mathbb{P}[C_1,C_2],
\end{align}
using \Cref{ass:iv2}.1 and \ref{ass:iv2}.3.

The first stage with $D=D_1D_2=D_{\land}$ uses that under \Cref{ass:double_exclusion}
\begin{align}
D_{\land}=&[D_1(1)Z+D_1(0)(1-Z)]\times [D_2(0)+(D_2(1)-D_2(0))(D_1(1)Z+D_1(0)(1-Z))] \\ \notag
       =&D_1(1)D_2(0)Z+(D_2(1)-D_2(0))D_1(1)Z+ \\ \notag
       &D_1(0)D_2(0)(1-Z)+(D_2(1)-D_2(0))D_1(0)(1-Z)\\ \notag
        =&D_1(1)D_2(1)Z+D_1(0)D_2(1)(1-Z),
\end{align}
so that the first stage equals 
\begin{align}\label{eq:fspo_dand_double}
 \Delta\mathbb{E}[D_{\land}|Z]=&\mathbb{E}[D_1(1)D_2(1)Z+D_1(0)D_2(1)(1-Z)|Z=1]-\\ \notag
    &\mathbb{E}[D_1(1)D_2(1)Z+D_1(0)D_2(1)(1-Z)|Z=0] \\ \notag
    =&\mathbb{E}[(D_1(1)-D_1(0))D_2(1)]=\mathbb{P}[(D_1(1)-D_1(0))D_2(1)=1]\\ \notag
     =& \mathbb{P}[D_1(1)-D_1(0)=1,D_2(1)-D_2(0)=1] +\mathbb{P}[D_1(1)-D_1(0)=1,D_2(1)=D_2(0)=1]\\ \notag
    =&\mathbb{P}[C_1,C_2]+\mathbb{P}[C_1,A_2],
\end{align}
using \Cref{ass:iv2}.1 and \ref{ass:iv2}.3.

The first stage with $D=D_\lor=D_1+D_2-D_1D_2$ equals
\begin{align}\label{eq:fspo_dor_double}
    \Delta\mathbb{E}[D_\lor|Z]=&\Delta\mathbb{E}[D_1|Z]+\Delta\mathbb{E}[D_2|Z]-\Delta\mathbb{E}[D_{\land}|Z] \\ \notag
    =& \mathbb{P}[C_1,C_2]+\mathbb{P}[C_1,N_2].
\end{align}

For the reduced form, the proof of \Cref{lemma:fs_rf} shows that
\begin{align}
     \Delta\mathbb{E}[Y|Z]
    =&\mathbb{E}[(Y(1,1)-Y(0,1)-Y(1,0)+Y(0,0))D_1D_2|Z=1]-  \\ \notag 
    &\mathbb{E}[(Y(1,1)-Y(0,1)-Y(1,0)+Y(0,0))D_1D_2|Z=0]+\\ \notag
    &\mathbb{E}[(Y(1,0)-Y(0,0))D_1|Z=1]-\mathbb{E}[(Y(1,0)-Y(0,0))D_1|Z=0]+ \\ \notag
    &\mathbb{E}[(Y(0,1)-Y(0,0))D_2|Z=1]-\mathbb{E}[(Y(0,1)-Y(0,0))D_2|Z=0],
\end{align}
using \Cref{ass:iv2}.4. It follows from the results above in \eqref{eq:fspo_d1}, \eqref{eq:fspo_d2_double}, and \eqref{eq:fspo_dand_double} that
\begin{align}
\Delta\mathbb{E}[Y|Z]
    =& \mathbb{E}[Y(1,1)-Y(0,1)-Y(1,0)+Y(0,0)|\{C_1,C_2\},\{C_1,A_2\}]\mathbb{P}[\{C_1,C_2\},\{C_1,A_2\}]+\\ \notag
    &\mathbb{E}[Y(1,0)-Y(0,0)|\{C_1,C_2\},\{C_1,N_2\},\{C_1,A_2\}]\mathbb{P}[\{C_1,C_2\},\{C_1,N_2\},\{C_1,A_2\}]+ \\  \notag
    &\mathbb{E}[Y(0,1)-Y(0,0)|C_1,C_2]\mathbb{P}[C_1,C_2]\\ \notag
    =& \mathbb{E}[Y(1,1)-Y(0,1)-Y(1,0)+Y(0,0)|C_1,C_2]\mathbb{P}[C_1,C_2]+\\  \notag
    &\mathbb{E}[Y(1,1)-Y(0,1)-Y(1,0)+Y(0,0)|C_1,A_2]\mathbb{P}[C_1,A_2]+ \\ \notag
    &\mathbb{E}[Y(1,0)-Y(0,0)|C_1,C_2]\mathbb{P}[C_1,C_2]+
    \mathbb{E}[Y(1,0)-Y(0,0)|C_1,N_2]\mathbb{P}[C_1,N_2]+\\  \notag
    &\mathbb{E}[Y(1,0)-Y(0,0)|C_1,A_2]\mathbb{P}[C_1,A_2]+\mathbb{E}[Y(0,1)-Y(0,0)|C_1,C_2]\mathbb{P}[C_1,C_2]\\ \notag
    =& \mathbb{E}[Y(1,1)-Y(0,0)|C_1,C_2]\mathbb{P}[C_1,C_2]+\\  \notag
    &\mathbb{E}[Y(1,0)-Y(0,0)|C_1,N_2]\mathbb{P}[C_1,N_2]+\mathbb{E}[Y(1,1)-Y(0,1)|C_1,A_2]\mathbb{P}[C_1,A_2].
\end{align}

\section{\texorpdfstring{Proof \Cref{theorem:late_c}}{proof}}\label{A:proof_theorem3}

First, we collect some relations that will be useful for this proof and the proof of \Cref{corr:bounded}. Note that from \Cref{lemma:fs_rf_double} we have the following relations:
\begin{align}
\Delta\mathbb{E}[D_{\lor}-D_2|Z]&= \Delta\mathbb{E}[D_1|Z]-\Delta\mathbb{E}[D_{\land}|Z] =\mathbb{P}[C_1,N_2].\\
\Delta\mathbb{E}[D_{\land}-D_2|Z]&= \Delta\mathbb{E}[D_{\land}|Z]-\Delta\mathbb{E}[D_2|Z]=\mathbb{P}[C_1,A_2].
\end{align}
Using \eqref{eq:Yd}, we can write
\begin{align}
    \label{eq:po_d1y}
    D_1Y&=[Y(1,1)-Y(1,0)]D_1D_2+Y(1,0)D_1, \\
    \label{eq:po_d2y}
    D_2Y&=[Y(1,1)-Y(0,1)]D_1D_2+Y(0,1)D_2, \\
    \label{eq:po_d1d2y}
    D_1D_2Y&=Y(1,1)D_1D_2,
\end{align}
and it follows from the results in \Cref{lemma:fs_rf_double} that
\begin{align}
\Delta\mathbb{E}[D_1Y|Z] =& \Delta\mathbb{E}[[Y(1,1)-Y(1,0)]D_1D_2|Z] + \Delta\mathbb{E}[Y(1,0)D_1|Z] \\ \notag
=& \mathbb{E}[(Y(1,1)-Y(1,0))D_1D_2|Z=1]-\mathbb{E}[(Y(1,1)-Y(1,0))D_1D_2|Z=0]+ \\ \notag
&\mathbb{E}[(Y(1,0)D_1|Z=1]-\mathbb{E}[(Y(1,0)D_1|Z=0] \\ \notag
=&\mathbb{E}[Y(1,1)-Y(1,0)|C_1,C_2]\mathbb{P}[C_1,C_2] + 
\mathbb{E}[Y(1,1)-Y(1,0)|C_1,A_2]\mathbb{P}[C_1,A_2] + \\ \notag
&\mathbb{E}[Y(1,0)|C_1,C_2]\mathbb{P}[C_1,C_2] + 
\mathbb{E}[Y(1,0)|C_1,N_2]\mathbb{P}[C_1,N_2] \\ \notag
&\mathbb{E}[Y(1,0)|C_1,A_2]\mathbb{P}[C_1,A_2] \\ \notag
=&\mathbb{E}[Y(1,1)|C_1,C_2]\mathbb{P}[C_1,C_2] + \mathbb{E}[Y(1,0)|C_1,N_2]\mathbb{P}[C_1,N_2]+\\ \notag
&\mathbb{E}[Y(1,1)|C_1,A_2]\mathbb{P}[C_1,A_2],
\end{align}
and
\begin{align}
    \Delta\mathbb{E}[D_2Y|Z] =& \Delta\mathbb{E}[[Y(1,1)-Y(0,1)]D_1D_2|Z] + \Delta\mathbb{E}[Y(0,1)D_2|Z] \\ \notag
    =& \mathbb{E}[(Y(1,1)-Y(0,1))D_1D_2|Z=1]-\mathbb{E}[(Y(1,1)-Y(0,1))D_1D_2|Z=0]+ \\ \notag
    &\mathbb{E}[Y(0,1)D_2|Z=1]-\mathbb{E}[Y(0,1)D_2|Z=0] \\ \notag 
    =&\mathbb{E}[Y(1,1)-Y(0,1)|C_1,C_2]\mathbb{P}[C_1,C_2] + 
    \mathbb{E}[Y(1,1)-Y(0,1)|C_1,A_2]\mathbb{P}[C_1,A_2] + \\ \notag
    &\mathbb{E}[Y(0,1)|C_1,C_2]\mathbb{P}[C_1,C_2] \\ \notag
    =&\mathbb{E}[Y(1,1)|C_1,C_2]\mathbb{P}[C_1,C_2] + \mathbb{E}[Y(1,1)-Y(0,1)|C_1,A_2]\mathbb{P}[C_1,A_2], 
\end{align}
and
\begin{align}
\Delta\mathbb{E}[D_{\land}Y|Z] =& \Delta\mathbb{E}[Y(1,1)D_1D_2|Z] \\ \notag
  =& \mathbb{E}[Y(1,1)D_1D_2|Z=1]-\mathbb{E}[Y(1,1)D_1D_2|Z=0] \\ \notag
    =&\mathbb{E}[Y(1,1)|C_1,C_2]\mathbb{P}[C_1,C_2] + \mathbb{E}[Y(1,1)|C_1,A_2]\mathbb{P}[C_1,A_2].
\end{align}

From this we can construct,
\begin{align}
\Delta\mathbb{E}[(D_{\lor}-D_2)Y|Z] =&  \Delta\mathbb{E}[D_1Y|Z]- \Delta\mathbb{E}[D_{\land}Y|Z] \\ \notag 
=& \mathbb{E}[Y(1,0)|C_1,N_2]\mathbb{P}[C_1,N_2],\\
\Delta\mathbb{E}[(D_{\land}-D_2)Y|Z]  =&   \Delta\mathbb{E}[D_{\land}Y|Z] -\Delta\mathbb{E}[D_2Y|Z]\\ \notag 
=& \mathbb{E}[Y(0,1)|C_1,A_2]\mathbb{P}[C_1,A_2], \\
\Delta\mathbb{E}[(1-D_1)(1-D_2)Y|Z]=& \Delta\mathbb{E}[Y|Z] + \Delta\mathbb{E}[D_{\land}Y|Z] -\Delta\mathbb{E}[D_1Y|Z] -\Delta\mathbb{E}[D_2Y|Z]\\  \notag
=&-\mathbb{E}[Y(0,0)|C_1,C_2]\mathbb{P}[C_1,C_2]-\mathbb{E}[Y(0,0)|C_1,N_2]\mathbb{P}[C_1,N_2].  
\end{align}

Second, we construct the lower and upper bounds. Assume the following monotone treatment response: $\mathbb{E}[Y(1,1)-Y(1,0)|C_1,N_2] \geq 0$ and $\mathbb{E}[Y(0,1)-Y(0,0)|C_1,A_2] \geq 0$. Using the result in \Cref{lemma:fs_rf_double}, we can construct the lower bound as
\begin{align}
    \Delta\mathbb{E}[Y|Z] =& \mathbb{E}[Y(1,1)-Y(0,0)|C_1,C_2]\mathbb{P}[C_1,C_2]+\\ \notag
    &\mathbb{E}[Y(1,0)-Y(0,0)|C_1,N_2]\mathbb{P}[C_1,N_2]+\mathbb{E}[Y(1,1)-Y(0,1)|C_1,A_2]\mathbb{P}[C_1,A_2]\\ \notag
    =& \mathbb{E}[Y(1,1)-Y(0,0)|C_1,C_2]\mathbb{P}[C_1,C_2]+\\ \notag
    &\mathbb{E}[Y(1,1)-Y(0,0)|C_1,N_2]\mathbb{P}[C_1,N_2]+\mathbb{E}[Y(1,1)-Y(0,0)|C_1,A_2]\mathbb{P}[C_1,A_2]-\\ \notag
    &\mathbb{E}[Y(1,1)-Y(1,0)|C_1,N_2]\mathbb{P}[C_1,N_2]-\mathbb{E}[Y(0,1)-Y(0,0)|C_1,A_2]\mathbb{P}[C_1,A_2]\\ \notag
    \leq&\mathbb{E}[Y(1,1)-Y(0,0)|\{C_1,C_2\},\{C_1,N_2\},\{C_1,A_2\}](\mathbb{P}[\{C_1,C_2\},\{C_1,N_2\},\{C_1,A_2\}].
\end{align}
And from \Cref{lemma:fs_rf_double} follows that $ \Delta\mathbb{E}[D_1|Z]=\mathbb{P}[C_1,C_2]+\mathbb{P}[C_1,N_2]+\mathbb{P}[C_1,A_2]$.

Assume the following monotone treatment selection: $\mathbb{E}[Y(1,1)|C_1,A_2] \geq \mathbb{E}[Y(1,1)|C_1,N_2]$ and $\mathbb{E}[Y(1,1)|C_1,C_2] \geq \mathbb{E}[Y(1,1)|C_1,N_2]$. Using the results above, it follows that
\begin{align}
    \frac{\Delta\mathbb{E}[D_\land Y|Z]}{\Delta\mathbb{E}[D_\land|Z]} =& \frac{\mathbb{E}[Y(1,1)|C_1,C_2]\mathbb{P}[C_1,C_2]+\mathbb{E}[Y(1,1)|C_1,A_2]\mathbb{P}[C_1,A_2]}{\mathbb{P}[C_1,C_2]+\mathbb{P}[C_1,A_2]}\\  \notag
    \geq&\frac{\mathbb{E}[Y(1,1)|C_1,C_2]\mathbb{P}[C_1,C_2]+\mathbb{E}[Y(1,1)|C_1,A_2]\mathbb{P}[C_1,A_2]}{\mathbb{P}[C_1,C_2]+\mathbb{P}[C_1,A_2]+\mathbb{P}[C_1,N_2]}+\\  \notag
    &\frac{\mathbb{E}[Y(1,1)|C_1,N_2]\mathbb{P}[C_1,N_2]}{\mathbb{P}[C_1,C_2]+\mathbb{P}[C_1,A_2]+\mathbb{P}[C_1,N_2]}.
\end{align}

Next, assume the following positive response: $\mathbb{E}[Y(0,0)|C_1,A_2]\geq 0$. Using the results above, it follows that
\begin{align}
    \frac{\Delta\mathbb{E}[(1-D_1)(1-D_2)Y|Z]}{\Delta\mathbb{E}[D_1|Z]}=&\frac{-\mathbb{E}[Y(0,0)|C_1,C_2]\mathbb{P}[C_1,C_2]-\mathbb{E}[Y(0,0)|C_1,N_2]\mathbb{P}[C_1,N_2]}{\mathbb{P}[C_1,C_2]+\mathbb{P}[C_1,N_2]+\mathbb{P}[C_1,A_2]}\\  \notag
    \geq&\frac{-\mathbb{E}[Y(0,0)|C_1,C_2]\mathbb{P}[C_1,C_2]-\mathbb{E}[Y(0,0)|C_1,N_2]\mathbb{P}[C_1,N_2]}{\mathbb{P}[C_1,C_2]+\mathbb{P}[C_1,N_2]+\mathbb{P}[C_1,A_2]}-\\  \notag
    &\frac{\mathbb{E}[Y(0,0)|C_1,A_2]\mathbb{P}[C_1,A_2]}{\mathbb{P}[C_1,C_2]+\mathbb{P}[C_1,N_2]+\mathbb{P}[C_1,A_2]}.
\end{align}
We can construct the upper bound as
\begin{align}
&\frac{\Delta\mathbb{E}[D_\land Y|Z]}{\Delta\mathbb{E}[D_\land|Z]}+\frac{\Delta\mathbb{E}[(1-D_1)(1-D_2)Y|Z]}{\Delta\mathbb{E}[D_1|Z]}\geq\\ \notag
&\frac{\mathbb{E}[Y(1,1)-Y(0,0)|C_1,C_2]\mathbb{P}[C_1,C_2]+\mathbb{E}[Y(1,1)-Y(0,0)|C_1,N_2]\mathbb{P}[C_1,N_2]}{\mathbb{P}[C_1,C_2]+\mathbb{P}[C_1,N_2]+\mathbb{P}[C_1,A_2]}+\\ \notag
&\frac{\mathbb{E}[Y(1,1)-Y(0,0)|C_1,A_2]\mathbb{P}[C_1,A_2]}{\mathbb{P}[C_1,C_2]+\mathbb{P}[C_1,N_2]+\mathbb{P}[C_1,A_2]}.
\end{align}

\section{\texorpdfstring{Proof \Cref{corr:bounded}}{proof}}\label{A:corr_bounded}

Using the results in \Cref{A:proof_theorem3}, construct the lower bound as
\begin{align}
    &\Delta\mathbb{E}[(1-D_1-D_2+2D_1D_2)Y|Z]+Y_{\min}\Delta\mathbb{E}[D_\lor-D_2|Z]-Y_{\max}\Delta\mathbb{E}[D_\land-D_2|Z]= \\ \notag
   & \Delta\mathbb{E}[Y|Z]+\Delta\mathbb{E}[(D_{\land}-D_2)Y|Z]-\Delta\mathbb{E}[(D_{\lor}-D_2)Y|Z]+\\ \notag
   &Y_{\min}\mathbb{P}[C_1,N_2]-Y_{\max}\mathbb{P}[C_1,A_2] =\\ \notag
    & \mathbb{E}[Y(1,1)-Y(0,0)|C_1,C_2]\mathbb{P}[C_1,C_2]+
    \mathbb{E}[Y_{\min}-Y(0,0)|C_1,N_2]\mathbb{P}[C_1,N_2]+\\ \notag
    &\mathbb{E}[Y(1,1)-Y_{\max}|C_1,A_2]\mathbb{P}[C_1,A_2]\leq \\ \notag
    &\mathbb{E}[Y(1,1)-Y(0,0)|\{C_1,C_2\},\{C_1,N_2\},\{C_1,A_2\}]\mathbb{P}[\{C_1,C_2\},\{C_1,N_2\},\{C_1,A_2\}].
\end{align}
Similarly, the upper bound equals
\begin{align}
    & \Delta\mathbb{E}[(1-D_1-D_2+2D_1D_2)Y|Z]+Y_{\max}\Delta\mathbb{E}[D_\lor-D_2|Z]-Y_{\min}\Delta\mathbb{E}[D_\land-D_2|Z]=\\\notag
   & \Delta\mathbb{E}[Y|Z]+\Delta\mathbb{E}[(D_{\land}-D_2)Y|Z]-\Delta\mathbb{E}[(D_{\lor}-D_2)Y|Z]+\\ \notag
   &Y_{\max}\mathbb{P}[C_1,N_2]-Y_{\min}\mathbb{P}[C_1,A_2] =\\ \notag
    & \mathbb{E}[Y(1,1)-Y(0,0)|C_1,C_2]\mathbb{P}[C_1,C_2]+
    \mathbb{E}[Y_{\max}-Y(0,0)|C_1,N_2]\mathbb{P}[C_1,N_2]+\\  \notag
    &\mathbb{E}[Y(1,1)-Y_{\min}|C_1,A_2]\mathbb{P}[C_1,A_2]\geq \\  \notag
    &\mathbb{E}[Y(1,1)-Y(0,0)|\{C_1,C_2\},\{C_1,N_2\},\{C_1,A_2\}]\mathbb{P}[\{C_1,C_2\},\{C_1,N_2\},\{C_1,A_2\}].
\end{align}
And from \Cref{lemma:fs_rf_double} follows that $\Delta\mathbb{E}[D_1|Z] = \mathbb{P}[C_1,C_2] + \mathbb{P}[C_1,N_2]+ \mathbb{P}[C_1,A_2] $.

\section{\texorpdfstring{Proof \Cref{prop:nec_cond_movers}}{proof}}\label{A:proof_prop1}
Note that we do not invoke \Cref{ass:double_exclusion} in this proof. Hence, we use the expressions in \Cref{lemma:fs_rf} to show that
\begin{align}
    \Delta\mathbb{E}[D_{\lor}-D_2|Z] =& \Delta\mathbb{E}[D_1|Z]-\Delta\mathbb{E}[D_{\land}|Z]=\mathbb{P}[C_1,N_2]-\mathbb{P}[A_1,C_2],\\
    \Delta\mathbb{E}[D_{\land}-D_2|Z] =& \Delta\mathbb{E}[D_{\land}|Z]-\Delta\mathbb{E}[D_2|Z]=\mathbb{P}[C_1,A_2]-\mathbb{P}[N_1,C_2].
\end{align}
Using \eqref{eq:po_d1y}, \eqref{eq:po_d2y}, and \eqref{eq:po_d1d2y},  
and using the results in \eqref{eq:fspo_d1} and \eqref{eq:fspo_dand}, we have
\begin{align}
    \Delta\mathbb{E}[D_1Y|Z]=&\mathbb{E}[(Y(1,1)-Y(1,0))D_1D_2+Y(1,0)D_1|Z=1]- \\ \notag
    &\mathbb{E}[(Y(1,1)-Y(1,0))D_1D_2+Y(1,0)D_1|Z=0] \\ \notag
    =&\mathbb{E}[(Y(1,1)-Y(1,0))|\{C_1,C_2\},\{C_1,A_2\},\{A_1,C_2\}]\mathbb{P}[\{C_1,C_2\},\{C_1,A_2\},\{A_1,C_2\}]+ \\ \notag
    &\mathbb{E}[Y(1,0)|\{C_1,C_2\},\{C_1,N_2\},\{C_1,A_2\}]\mathbb{P}[\{C_1,C_2\},\{C_1,N_2\},\{C_1,A_2\}] \\ \notag
    =& \mathbb{E}[Y(1,1)|C_1,C_2]\mathbb{P}[C_1,C_2]+\mathbb{E}[Y(1,0)|C_1,N_2]\mathbb{P}[C_1,N_2]+\\ \notag
    &\mathbb{E}[Y(1,1)|C_1,A_2]\mathbb{P}[C_1,A_2]+\mathbb{E}[Y(1,1)-Y(1,0)|A_1,C_2]\mathbb{P}[A_1,C_2].
\end{align}
If we additionally use \eqref{eq:fspo_d2}, we similarly have that
\begin{align}
    \Delta\mathbb{E}[D_2Y|Z] =&\mathbb{E}[Y(1,1)|C_1,C_2]\mathbb{P}[C_1,C_2]+\mathbb{E}[Y(0,1)|N_1,C_2]\mathbb{P}[N_1,C_2]+\\ \notag
    &\mathbb{E}[Y(1,1)|A_1,C_2]\mathbb{P}[A_1,C_2]+\mathbb{E}[Y(1,1)-Y(0,1)|C_1,A_2]\mathbb{P}[C_1,A_2],\\
    \Delta\mathbb{E}[D_{\land}Y|Z] =& \mathbb{E}[Y(1,1)|C_1,C_2]\mathbb{P}[C_1,C_2]+\mathbb{E}[Y(1,1)|C_1,A_2]\mathbb{P}[C_1,A_2]+\\  \notag
    &\mathbb{E}[Y(1,1)|A_1,C_2]\mathbb{P}[A_1,C_2],
\end{align}
from which follows that
\begin{align}
    \Delta\mathbb{E}[(D_{\lor}-D_2)Y|Z] =& \Delta\mathbb{E}[D_1Y|Z]-\Delta\mathbb{E}[D_{\land}Y|Z]\\ \notag
    =& \mathbb{E}[Y(1,0)|C_1,N_2]\mathbb{P}[C_1,N_2]-\mathbb{E}[Y(1,0)|A_1,C_2]\mathbb{P}[A_1,C_2],\\
    \Delta\mathbb{E}[(D_{\land}-D_2)Y|Z] =& \Delta\mathbb{E}[D_{\land}Y|Z]-\Delta\mathbb{E}[D_2Y|Z]\\ \notag
    =& \mathbb{E}[Y(0,1)|C_1,A_2]\mathbb{P}[C_1,A_2]-\mathbb{E}[Y(0,1)|N_1,C_2]\mathbb{P}[N_1,C_2].
\end{align}

\section{\texorpdfstring{Proof \Cref{corr:partialIV}}{proof}}\label{A:proof_corr_partial}

\Cref{corr:partialIV} defines the weighted average of LATEs
\begin{align} 
    \tau = &\mathbb{E}[Y(1,1)-Y(0,0)|C_1,C_2]w[C_1,C_2]+ 
    \mathbb{E}[Y(1,0)-Y(0,0)|C_1,N_2]w[C_1,N_2]+ \\ \notag
    &\,\mathbb{E}[Y(1,1)-Y(0,1)|C_1,A_2]w[C_1,A_2]+ 
    \mathbb{E}[Y(0,1)-Y(0,0)|N_1,C_2]w[N_1,C_2] +\\ \notag
     &\, \mathbb{E}[Y(1,1)-Y(1,0)|A_1,C_2]w[A_1,C_2],
\end{align}
with $w(G_1,G_2)=\frac{\mathbb{P}[G_1,G_2]}{\mathbb{P}[C_1,C_2]+\mathbb{P}[C_1,N_2]+\mathbb{P}[C_1,A_2]+\mathbb{P}[N_1,C_2]+\mathbb{P}[A_1,C_2]}$. From \Cref{lemma:fs_rf} follows that
\begin{align}
   \Delta\mathbb{E}[Y|Z]=\tau (\mathbb{P}[C_1,C_2]+\mathbb{P}[C_1,N_2]+\mathbb{P}[C_1,A_2]+\mathbb{P}[N_1,C_2]+\mathbb{P}[A_1,C_2]).
\end{align}
From this we can construct an upper bound on $\tau$:
\begin{align}
\beta_{IV}(D_1)=\frac{\Delta\mathbb{E}[Y|Z]}{\Delta\mathbb{E}[D_1|Z]}=&\tau \frac{\mathbb{P}[C_1,C_2]+\mathbb{P}[C_1,N_2]+\mathbb{P}[C_1,A_2]+\mathbb{P}[N_1,C_2]+\mathbb{P}[A_1,C_2]}{\mathbb{P}[C_1,C_2]+\mathbb{P}[C_1,N_2]+\mathbb{P}[C_1,A_2]}\\ \notag
=&\tau \Big(1+ \frac{\mathbb{P}[N_1,C_2]+\mathbb{P}[A_1,C_2]}{\mathbb{P}[C_1,C_2]+\mathbb{P}[C_1,N_2]+\mathbb{P}[C_1,A_2]}\Big)
\geq \tau.
\end{align}
However, if other specifications for $D$ give a smaller second stage estimate, this is a tighter upper bound. 

From \Cref{lemma:fs_rf} also follows that
\begin{align}
    \Delta\mathbb{E}[D_1+D_2|Z]=&\Delta\mathbb{E}[D_1|Z]+\Delta\mathbb{E}[D_2|Z] \\ \notag
    =& 2\mathbb{P}[C_1,C_2]+\mathbb{P}[C_1,N_2]+\mathbb{P}[C_1,A_2]+\mathbb{P}[N_1,C_2]+\mathbb{P}[A_1,C_2].
\end{align}
From this we can construct a lower bound on $\tau$:
\begin{align}
\beta_{IV}(D_1+D_2)=&\frac{\Delta\mathbb{E}[Y|Z]}{\Delta\mathbb{E}[D_1+D_2|Z]}\\ \notag
=&\tau \frac{\mathbb{P}[C_1,C_2]+\mathbb{P}[C_1,N_2]+\mathbb{P}[C_1,A_2]+\mathbb{P}[N_1,C_2]+\mathbb{P}[A_1,C_2]}{2\mathbb{P}[C_1,C_2]+\mathbb{P}[C_1,N_2]+\mathbb{P}[C_1,A_2]+\mathbb{P}[N_1,C_2]+\mathbb{P}[A_1,C_2]}\\ \notag
=&\tau \Big(1 - \frac{\mathbb{P}[C_1,C_2]}{2\mathbb{P}[C_1,C_2]+\mathbb{P}[C_1,N_2]+\mathbb{P}[C_1,A_2]+\mathbb{P}[N_1,C_2]+\mathbb{P}[A_1,C_2]}\Big)
\leq \tau.
\end{align}

\section{Additional details on the empirical specifications}\label{A:app}

\subsection{The Oregon health insurance experiment}

Similar to \cite{finkelstein2012oregon, finkelstein2016effect}, the outcome variable $Y$ is measured during the same period as the treatment variables $D_1$ and $D_2$. For IV estimation to be valid in this specification, emergency department use should not affect Medicaid coverage. We find qualitatively similar results if we measure $D_1$ in the first year, $D_2$ in the first half of the second year, and $Y$ in the second year after treatment.

We follow the specification of the controls and standard errors in \cite{finkelstein2016effect}. We include three control variables in our analysis. First, since the entire household of any selected individual had the opportunity to apply for Medicaid, we include household size dummies. Second, since the lottery draws were done in eight rounds, we include lottery round dummies. Third, to improve precision, we control for the outcome variable $Y$ measured during the 14 months prior to the first lottery draw. Standard errors are clustered at the household level.

\subsection{The LinkedIn job opportunities experiment}

LinkedIn usage is measured via extractions from the Linkedin administrative data at the end of the job training program and again six (and twelve) months later. Employment six and twelve months after the job training program is measured via phone surveys. The response to the phone survey is, respectively, 68\% and 60\% six and twelve months after the program ended. This response rate is balanced across the treatment and control group. The final estimation sample is restricted to the 988 participants for whom we observe employment status twelve months after the program. 

The random assignment took place on the cohort level. In particular, the 30 cohorts were randomly split into 15 treatment and 15 control groups using within-city, sequentially paired randomization. We follow the specification of the controls and standard errors in  \cite{wheeler2022linkedin}. We include cohort-pair dummies as control variables. Standard errors are clustered at the cohort level.

\subsection{The elite university education natural experiment}

The application score equally weighs a combination of high school grades and an admission test score administered by the elite university. Due to the short time span between the admission test and enrollment deadline, the elite university does not have a waiting list but sets the number of admitted students slightly above capacity based upon previous years' enrollment rates.

\cite{anelli2020returns} compares applicants just above and below the application score cutoffs via a fuzzy Regression Discontinuity Design (RDD). The number of students, pre-treatment characteristics, and the probability of observing zero or missing income are smooth across the cutoff. The sample minimum and maximum of log income ($Y$) is equal to $2.398$ and $12.441$ respectively.

The polynomial and bandwidth are important considerations for an RD analysis. \cite{anelli2020returns} uses a polynomial of degree 0, which is often referred to as the local randomization approach \citep{cattaneo2023practical}, and uses four different bandwidths: no bandwidth using all 645 applicants (Table 4, column (2) of \cite{anelli2020returns}), the MSE-optimal bandwidth using 162 to 175 applicants depending on the specification (Table 4, column (3)), and twice and three times the MSE-optimal bandwidth using 280 to 303 and 382 to 409 applicants respectively (Table 4, column (4) and (5)). Results are qualitatively similar across the four bandwidths. 

Our results are based on the specification without bandwidth. The RD specification, absent polynomial and bandwidth, follows a standard IV setup with a binary instrument $Z$ that equals one if the applicant had a score above the cutoff. We follow the specification of the controls and standard errors in \cite{anelli2020returns}. We include five control variables: parental house value, a female dummy, a dummy for students commuting from outside the city, fixed effects for the year of application, and a (potentially endogenous) dummy for whether the individual is enrolled in university in 2005, which is the year when income is observed. Standard errors are clustered at the cohort high-school level.

Ideally our $D_2$ would measure whether the student graduated from the elite university. In this case, the bounds would partially identify the LAFTE of the elite university. Since students that were induced to enroll at the elite university may never graduate from this institution, it is still likely that $\mathbb{P}[C_1,N_2]>0$. In turn, with this $D_2$ it may be that $\mathbb{P}[C_1,A_2]=\mathbb{P}[N_1,C_2]=\mathbb{P}[A_1,C_2]=0$. The first two mover types would likely be absent since a student cannot graduate from the elite university without ever being enrolled. The third mover type would be absent since there are very few always takers with elite enrollment. 

\subsection{The village-based schools experiment}

The village-based schools were opened in the summer of 2007. The researchers conducted surveys of all available households in the control and treatment villages in the fall of 2007, four months after the opening, and in the spring of 2008, eight months after the opening. Both surveys determined school enrollment status of each school-age child living in the household and administered a short test covering math and language. Survey coverage rates were similar across treatment and control villages.

We follow \cite{burde2013bringing} and exclude a small number of extremely large and wealthy households and further restrict the estimation sample to the children for whom we observe enrollment in both surveys and test scores in the spring 2008 survey. The final sample includes 1,181 children, of in total 1,490 school-age children across the 31 villages. The sample minimum and maximum of the standardized test scores in the spring 2008 survey ($Y$) is equal to $-1.263$ and $2.662$ respectively. 

We follow the specification of the controls and standard errors in \cite{burde2013bringing}. We  control for a single dummy variable that equals one if the village is located in the Chagcharan district. Since the 31 villages were grouped into 11 equally sized village groups, randomization took place at the group level, and standard errors are clustered at the village-group level. 

\subsection{Sensitivity analysis controls}

The empirical results in this paper are based on the control specification used in the original papers. Here we analyse the robustness of our results against excluding the controls. We find in all four empirical applications identical mover types and identical results for the necessary conditions of the double exclusion restriction. The bounds of \Cref{theorem:late_c} are qualitatively similar except for inflated standard errors in all applications, and a lower bound close to zero in the Oregon health insurance experiment.

\section{Additional empirical results}\label{A:app2}

Column (1) and (2) of \Cref{tbl:bounds} report the precise estimates and standard errors of the bounds from \Cref{theorem:late_c}, which are visually reported in \Cref{fig:bounds}. Column (3) and (4) show the bounds from \Cref{corr:bounded} based on the bounded response. The latter two columns show that, for all four empirical applications, bounds based on the bounded response assumption are wide and cannot reject that the LAFTEs are equal to zero.

Table \ref{tbl:ss} reports the first stage and second stage estimates corresponding to different treatment variables. Column (1) shows the results for $D_1+D_2$, corresponding to the IV estimand in \eqref{eq:acr}. Columns (2) to (5) show the results for the treatment indicators $D_1$, $D_2$, $D_{\land}$, and $D_{\lor}$, respectively. Note that the estimate in column (1) does not require the double exclusion restriction and equals the lower bound for $\tau$ from \Cref{corr:partialIV}. The upper bound for $\tau$ equals the smallest second stage IV estimate in column (2) to (5). 
 
The double exclusion restriction is rejected for the village-based schools experiment in panel D. Column (1) allows us to conclude, however, that four months of education increases test scores by 0.679 standard deviations for the compliers. Subsequently, using the smallest second stage estimate in column (2) to (5), we can conclude that the estimated bounds for $\tau$ are $[0.679,1.301]$.

\begin{table}[t]
  \begin{center}
  \caption{Estimated bounds on the LAFTE} 
  \label{tbl:bounds}
    \begin{tabular}{ll*{4}{c}}
    \toprule \toprule
          &       & \multicolumn{2}{c}{Theorem 1 bounds} & \multicolumn{2}{c}{Corollary 3 bounds} \\
         &       & (1) & (2) & (3) & (4) \\
          &       & lower & upper & lower & upper \\
          \midrule
    \multicolumn{1}{l}{Panel A: Health insurance} & Estimate & 0.081 & 0.224 & -0.170 & 0.375 \\
          & Std. Error. & 0.024 & 0.026 & 0.025 & 0.023 \\ 
          \midrule
    \multicolumn{1}{l}{Panel B: LinkedIn training} & Estimate & 0.176 & 0.235 & 0.061 & 0.260 \\
          & Std. Error. & 0.066 & 0.078 & 0.086 & 0.082 \\ 
          \midrule
    \multicolumn{1}{l}{Panel C: Elite university} & Estimate & 0.532 & 7.545 & -2.772 & 5.865 \\
          & Std. Error. & 0.233 & 0.760  & 0.337 & 0.580 \\
          \bottomrule \bottomrule
    \end{tabular}
            \begin{tablenotes}
	\footnotesize
	\item[] Notes: this table shows the estimated lower and upper bounds from \Cref{theorem:late_c} and \Cref{corr:bounded} with their corresponding standard errors, for the empirical applications in \Cref{sec:application} for which the double exclusion restriction cannot be rejected. The upper bound from \Cref{theorem:late_c} is a linear function of two coefficients from two different second stage IV regressions. To obtain the standard error for this upper bound, we use a stacking approach that duplicates the data and estimates the two second stage IV coefficients via a single second stage IV regression. We cluster standard errors on the cluster variable that is similarly duplicated. 
\end{tablenotes}
\end{center}
\end{table}

\begin{table}[ht!]
\small
  \begin{center}
    \caption{The first and second stage estimates corresponding to different treatment variables}
    \label{tbl:ss}
\begin{tabular}{l*{5}{c}}
\toprule\toprule
                    &\multicolumn{1}{c}{(1)}&\multicolumn{1}{c}{(2)}&\multicolumn{1}{c}{(3)}&\multicolumn{1}{c}{(4)}&\multicolumn{1}{c}{(5)}\\
                    &\multicolumn{1}{c}{ $ D_{1}+D_{2} $ }&\multicolumn{1}{c}{ $ D_{1} $ }&\multicolumn{1}{c}{ $ D_{2} $ }&\multicolumn{1}{c}{ $ D_{\land} $ }&\multicolumn{1}{c}{ $ D_{\lor} $ }\\
                    \midrule
                     \multicolumn{6}{c}{Panel A: The Oregon health insurance experiment, $N=24646$} \\
                    & \multicolumn{5}{c}{First stage} \\
 $ Z $              &       0.369&       0.253&       0.115&       0.176&       0.193\\
                    &     (0.011)&     (0.006)&     (0.006)&     (0.006)&     (0.006)\\ 
                    [.25em]
                    & \multicolumn{5}{c}{Second stage} \\
$D$ from top row     &       0.056&   0.081&        0.179&        0.117&      0.107 \\
                    &     (0.016)&     (0.024)&      (0.052) &      (0.034)&    (0.031) \\
                    [.25em]
                    \midrule
                     \multicolumn{6}{c}{Panel B: The LinkedIn job opportunities experiment, $N=988$} \\
                    & \multicolumn{5}{c}{First stage} \\
 $ Z $              &       0.707&       0.393&       0.315&       0.357&       0.351\\
                    &     (0.148)&     (0.076)&     (0.077)&     (0.079)&     (0.074)\\
                    [.25em]
                    & \multicolumn{5}{c}{Second stage} \\
$D$ from top row   &       0.098&     0.176&      0.220&       0.194 &    0.197 \\
                    &     (0.039)&      (0.066)&   (0.096)&     (0.079) &     (0.077) \\
                    [.25em]
                    \midrule
                 \multicolumn{6}{c}{Panel C: The elite university education natural experiment, $N=645$} \\
                & \multicolumn{5}{c}{First stage} \\
$Z$                &       0.775&       0.680&       0.095&       0.633&       0.143\\
                    &     (0.106)&     (0.053)&     (0.061)&     (0.051)&     (0.063)\\
                    [.25em]
                & \multicolumn{5}{c}{Second stage} \\
$D$ from top row    &       0.466&     0.532&       3.796&      0.571 &       2.534  \\
                    &     (0.215)&     (0.233)&      (3.079)&       (0.254)&     (1.605)\\
                    [.25em]
                    \midrule
                    \multicolumn{6}{c}{Panel D: The village-based schools experiment , $N=1181$} \\
                    & \multicolumn{5}{c}{First stage} \\
 $ Z $              &       0.926&       0.443&       0.483&       0.443&       0.483\\
                    &     (0.158)&     (0.078)&     (0.080)&     (0.078)&     (0.080)\\
                    [.25em]
                    & \multicolumn{5}{c}{Second stage} \\
$D$ from top row     &       0.679&       1.420&      1.301&         1.420&    1.301 \\
                    &     (0.090)&      (0.191)&        (0.172)&      (0.191)&   (0.172)\\
                    [.25em]
       \bottomrule \bottomrule
    \end{tabular}
        \begin{tablenotes}
	\footnotesize
        \item[] Notes: this table provides the first and second stage estimates corresponding to different treatment variables. The numbers represent the estimated coefficients and the numbers in parentheses the corresponding standard errors. The panels correspond to the empirical applications in \Cref{sec:application}, where $N$ indicates the number of observations.
        \end{tablenotes}
 \end{center}
\end{table}

\end{document}